\documentclass[twocolumn]{aastex631}

\usepackage{pifont}
\newcommand{\cmark}{\ding{51}}%
\usepackage{threeparttable, tablefootnote}
\usepackage[version=3]{mhchem}

\begin{document}

\title{An ALMA molecular inventory of warm Herbig Ae disks: 
\\ I. Molecular rings, asymmetries and complexity in the HD~100546 disk}

\correspondingauthor{Alice S. Booth} 
\email{alice.booth@cfa.harvard.edu}

\author[0000-0003-2014-2121]{Alice S. Booth} 
\altaffiliation{Clay Postdoctoral Fellow}
\affiliation{Leiden Observatory, Leiden University, 2300 RA Leiden, the Netherlands}
\affiliation{Center for Astrophysics \textbar\, Harvard \& Smithsonian, 60 Garden St., Cambridge, MA 02138, USA}

\author[0000-0003-3674-7512]{Margot Leemker}
\affiliation{Leiden Observatory, Leiden University, 2300 RA Leiden, the Netherlands}

\author[0000-0001-7591-1907]{Ewine F. van Dishoeck}
\affiliation{Leiden Observatory, Leiden University, 2300 RA Leiden, the Netherlands}
\affiliation{Max-Planck-Institut für Extraterrestrische Physik, Giessenbachstrasse 1, 85748 Garching, Germany}

\author[0009-0006-1929-3896]{Lucy Evans}
\affiliation{School of Physics and Astronomy, University of Leeds, Leeds LS2 9JT, UK}

\author[0000-0003-1008-1142]{John D. Ilee}
\affiliation{School of Physics and Astronomy, University of Leeds, Leeds LS2 9JT, UK}

\author[0000-0003-0065-7267]{Mihkel Kama}
\affiliation{Department of Physics and Astronomy, University College London, Gower Street, London, WC1E 6BT, UK}
\affiliation{Tartu Observatory, University of Tartu, Observatooriumi 1, 61602 T\~{o}ravere, Tartumaa, Estonia}

\author[0000-0001-5849-577X]{Luke Keyte}
\affiliation{Department of Physics and Astronomy, University College London, Gower Street, London, WC1E 6BT, UK}

\author[0000-0003-1413-1776]{Charles J. Law}
\altaffiliation{NASA Hubble Fellowship Program Sagan Fellow}
\affiliation{Department of Astronomy, University of Virginia, Charlottesville, VA 22904, USA}

\author[0000-0003-2458-9756]{Nienke van der Marel}
\affiliation{Leiden Observatory, Leiden University, 2300 RA Leiden, the Netherlands}

\author[0000-0002-7058-7682]{Hideko Nomura}
\affiliation{Division of Science, National Astronomical Observatory of Japan, 2-21-1 Osawa, Mitaka, Tokyo 181-8588, Japan}

\author[0000-0003-2493-912X]{Shota Notsu}
\affiliation{Department of Earth and Planetary Science, Graduate School of Science, The University of Tokyo, 7-3-1 Hongo, Bunkyo-ku, Tokyo 113-0033, Japan}
\affiliation{Department of Astronomy, Graduate School of Science, The University of Tokyo, 7-3-1 Hongo, Bunkyo-ku, Tokyo 113-0033, Japan}
\affiliation{Star and Planet Formation Laboratory, RIKEN Cluster for Pioneering Research, 2-1 Hirosawa, Wako, Saitama 351-0198, Japan}

\author[0000-0001-8798-1347]{Karin Öberg}
\affiliation{Center for Astrophysics \textbar\, Harvard \& Smithsonian, 60 Garden St., Cambridge, MA 02138, USA}

\author[0000-0002-7935-7445]{Milou Temmink}
\affiliation{Leiden Observatory, Leiden University, 2300 RA Leiden, the Netherlands}

\author[0000-0001-6078-786X]{Catherine Walsh}
\affiliation{School of Physics and Astronomy, University of Leeds, Leeds LS2 9JT, UK}

\begin{abstract}
Observations of disks with the Atacama Large Millimeter/submillimeter Array (ALMA) allow us to map the chemical makeup of nearby protoplanetary disks with unprecedented spatial resolution and sensitivity. The typical outer Class II disk observed with ALMA is one with an elevated C/O ratio and a lack of oxygen-bearing complex organic molecules, but there are now some interesting exceptions: three transition disks around Herbig Ae stars all show oxygen-rich gas traced via the unique detections of the molecules SO and \ce{CH_3OH}. We present the first results of an ALMA line survey at $\approx$337 to 357~GHz of such disks and focus this paper on the first Herbig Ae disk to exhibit this chemical signature - HD~100546. In these data, we detect 19 different molecules including NO, \ce{SO_2} and \ce{CH_3OCHO} (methyl formate). We also make the first tentative detections of \ce{H_2^{13}CO} and \ce{^{34}SO} in protoplanetary disks. Multiple molecular species are detected in rings, which are, surprisingly, all peaking just beyond the underlying millimeter continuum ring at $\approx$200~au. This result demonstrates a clear connection between the large dust distribution and the chemistry in this flat disk. We discuss the physical and/or chemical origin of these sub-structures in relation to ongoing planet formation in the HD~100546 disk. We also investigate how similar and/or different the molecular make up of this disk is to other chemically well-characterised Herbig Ae disks. The line-rich data we present motivates the need for more ALMA line surveys to probe the observable chemistry in Herbig Ae systems which offer unique insight into the composition of disk ices, including complex organic molecules.
\end{abstract}

\section{Introduction} \label{sec:intro} 

There is now clear evidence that planet-formation is well underway in the $>$1~Myr old disks of gas, dust and ice around young stars \citep[e.g.][]{2018A&A...617A..44K, 2019NatAs...3..749H,2019ApJ...879L..25I}. Although the direct detection of giant planets in disks is still rare \citep{2018A&A...617A..44K} there are now a number of candidate planets \citep{2022arXiv220505696C, 2023MNRAS.522L..51H,2023NatAs...7.1208W}, and, the expected impact forming planets have on their parent disk is traceable in the millimetre dust emission, CO gas kinematics, and tentatively, in the disk chemistry \citep[e.g.,][]{Andrews2018dsharp, 2018ApJ...860L..12T, 2019NatAs...3.1109P, 2022ApJ...928....2I, 2022ApJ...937L...1L, 2022ApJ...934L..20B, 2023A&A...669A..53B, 2023arXiv230613710L}. The physical processes that drive planet formation and determine the composition of these forming planets are set by the parent protoplanetary disk \citep{2021PhR...893....1O, 2022arXiv220309818M}. Therefore, unravelling the physical and chemical conditions in disks is key to understanding the planet formation process. 

From both small population studies and detailed studies on a handful of individual disks with the Atacama Large Millimeter/submillimeter Array (ALMA) a picture of the ``typical" Class II disk composition has emerged. In most cases, the disk molecular layer on 10-100~au scales has an elevated carbon-to-oxygen ratio (C/O). This is traced via column densities of \ce{C_2H} that are orders of magnitude higher than initially predicted \citep[e.g.,][]{2019A&A...631A..69M}; disk averaged column density ratios of CS-to-SO that are greater than 1 \citep[e.g.,][]{2021ApJS..257...12L}; and complex molecules that lack oxygen (\ce{CH_3CN}, \ce{H_3CN} and \ce{c-C_3H_2}) are more abundant than anticipated \citep[e.g.,][]{2021ApJS..257....9I}. These results are all consistent with the gas-phase depletion of both oxygen and carbon in the molecular layer but with O more strongly depleted than C resulting in C/O$>$1 \citep{2016A&A...592A..83K, 2021ApJS..257....7B}.  This can be achieved by the removal of \ce{H_2O} and \ce{CO} from the disk molecular layer due to the growth, settling and drift of small ice-coated dust grains coupled with gas-phase and ice-phase chemical depletion mechanisms \citep{2015ApJ...807L..32D, 2018ApJ...856...85S, 2018A&A...618A.182B, 2020ApJ...899..134K, 2022A&A...665A..45P, 2022ApJ...938...29F, 2023NatAs...7...49C}. The depletion of gas-phase C and O implies that the ice is enriched in these elements. 

The above picture holds for disks around T-Tauri and Herbig Ae stars, although, for the latter, there are only a few sources for which the chemistry has been studied in detail. This includes the HD~163296 and MWC~480 disks, which we note are gapped disks but not transition disks with a large central cavity \citep[][]{2021ApJS..257....1O}. There are expected differences in the chemistry of these typically higher-mass sources due to the increased temperature of the host star and the different incident UV and X-ray radiation fields. Physical and chemical models predict warmer disks, lower levels of ionisation and, due to the increased FUV flux,
the warm molecular layer is expected to reside closer to the disk midplane \citep{2015A&A...582A..88W, 2018A&A...616A..19A}. Although the sample size for the Herbig Ae disks is small there is observational evidence to support these trends: this includes lower \ce{HCO^+}/CO column density ratios, less abundant \ce{H_2CO}, and lower detection rates of the cold chemistry tracers \ce{DCO^+} and DCN \citep{2004A&A...425..955T, 2020ApJ...893..101L, 2020ApJ...890..142P, 2021ApJS..257...13A, 2023arXiv230302167P, Booth2023}. 

Transition disks around young Herbig Ae type stars appear to be outliers from the ``typical" disk. In the few of these sources that have been observed in detail in molecular lines - HD 100546, IRS~48 and HD~169142 - a reservoir of oxygen-rich gas has been revealed primarily by emission lines of \ce{SO} and \ce{CH_3OH}.
These sources give us a window into the typically unobservable volatile reservoir in disks. 
First, SO and \ce{CH_3OH} were detected in the HD~100546 disk along with a compact component of \ce{H_2CO}, all of which thought to originate from the warm edge of the millimetre dust cavity \citep{2018A&A...611A..16B,2021NatAs...5..684B, 2023A&A...669A..53B}. Observations of IRS~48 then revealed the first detections of \ce{SO_2}, \ce{NO} and \ce{CH_3OCH_3} in protoplanetary disks \citep{2021A&A...651L...5V, 2021A&A...651L...6B, 2022A&A...659A..29B, 2023arXiv230300768L}. In addition, the measured rotational temperatures for \ce{H_2CO} and \ce{CH_3OH} in IRS~48 of $\approx$100-200~K thus indicate the conditions for \ce{H_2O} and COMs ice sublimation \citep{2021A&A...651L...5V}. Most recently, the HD~169142 disk was also revealed to have a reservoir of \ce{SO} and \ce{CH_3OH} within the millimetre dust cavity \citep{2023arXiv230613710L, Booth2023}. Although the observations are mostly spatially unresolved, \citet{Booth2023} report additional detections of DCN, CS, \ce{H_2CS}, \ce{HC_3N} and \ce{c-C_3H_2} - hinting at similarities in the chemistry of the HD~169142 disk to the non-transitional Herbig Ae disks \citep[e.g.,][]{2019ApJ...876...72L, 2021ApJS..257...12L, 2021ApJS..257....9I}. 
The only other chemically well studied transition disk around a Herbig source is HD~142527.  Observations so far do not show bright \ce{CH_3OH} or \ce{SO} emission, however, the millimetre dust (and ice) trap, is located further from the star where it remains too cold for thermal ice sublimation \citep{2023A&A...675A.131T}.

It is not yet clear how the chemistry in these warm Herbig Ae transition disks compares to other well-studied sources, both full Herbig Ae disks and T-Tauri disks. This is in terms of the abundances of both simple molecules, e.g., \ce{HCO^+}, \ce{HCN}, \ce{CN} and \ce{C_2H}, and, the more complex molecules, e.g., \ce{CH_3OH} and \ce{CH_3CN}, and, the spatial distribution of these species with respect to millimetre dust substructures. This paper presents ALMA Band 7 observations of the planet-forming disk HD~100546 where we target $>$20 molecules at $\approx$0\farcs3 ($\approx$30~au) resolution. These line-rich data enable us to put the HD~100546 disk in proper context with other Class II disks whose chemistry has been studied in greater detail. 
In paper II, \citet{Booth2023_irs48}, we present data on the IRS~48 disk which was observed in the same ALMA program. 
With this molecular inventory we determine the relationship (if any) between the different molecular tracers and the millimetre dust sub-structures, we use the simple molecules to constrain the underlying physical/chemical conditions, and we investigate the degree of molecular complexity that has been attained in this disk.

\section{Methods} \label{sec:methods}

\subsection{Target}
HD~100546 is nearby young Herbig Ae star host to a gas-rich protoplanetary disk (see Table~\ref{tab:A1}). The millimetre dust emission from the HD~100546 disk has been well studied with ALMA, which detected two dust rings peaking at $\approx$25 and $\approx$200~au \citep{2014ApJ...791L...6W, 2019ApJ...871...48P, 2021A&A...651A..90F}. Molecular line observations of the HD~100546 disk with ALMA have reported detections of the CO isotopologues \ce{^{12}CO}, \ce{^{13}CO}, \ce{C^{18}O} and the aforementioned detections of \ce{SO}, \ce{H_2CO} and \ce{CH_3OH} \citep{2019MNRAS.485..739M, 2018A&A...611A..16B, 2021NatAs...5..684B, 2023A&A...669A..53B}. Additionally, \ce{HCO^+} and \ce{CS} have been detected with the Australia Telescope Compact Array (ATCA) and the Atacama Compact Array (ACA), respectively \citep{Wright2015, 2023NatAs.tmp...92K}.

\begin{table*}
    \small{
    \centering
    \caption{Properties of the HD~100546 star disk system.}
    \begin{tabular}{c c c c c c c c c c c c} \hline \hline 
         SpT   & Dist. & Incl. & PA  & L$_{*}$ & M$_{*}$ & M$_{dust}$ & M$_{gas}$ & log$_{10}(\dot{M}_{acc}$) & log10($L_{Xray}$) & v$_{sys}$\\  
        &   (pc) & (deg) & (deg) & ($L_{\odot}$) &  ($M_{\odot}$) & ($M_{\odot}$) &($M_{\odot}$) &  ($M_{\odot}$ yr$^{-1}$) & (erg~s$^{-1}$) & (km s$^{-1}$)  \\  
        \hline 
          A0-A1 [1]  & 110 [2]  & 41.7 [3] & 146.0 [3] & 23.5 [2]  & 2.2 [2]& 1.1$\times10^{-3}$ [4] & 1.5$\times10^{-1}$ [4] &  -6.81 [1] & 28.1 [5] &  5.70 [3] \\ 
          \hline 
    \end{tabular}    
    \tablecomments{
    References: 
    [1]~\citet{2021A&A...650A.182G}, 
    [2]~\citet{2018A&A...620A.128V}, 
    [3]~\citet{2017A&A...607A.114W}, 
    [4]~\citet{Kama2016twhya}, 
    [5]~\citet{2012A&A...544A..78M}
    }
    }
    \label{tab:A1}
\end{table*}

\begin{figure*}
    \includegraphics[trim={3.25cm 0.75cm 3.25cm 1.25cm},clip,width=\hsize]{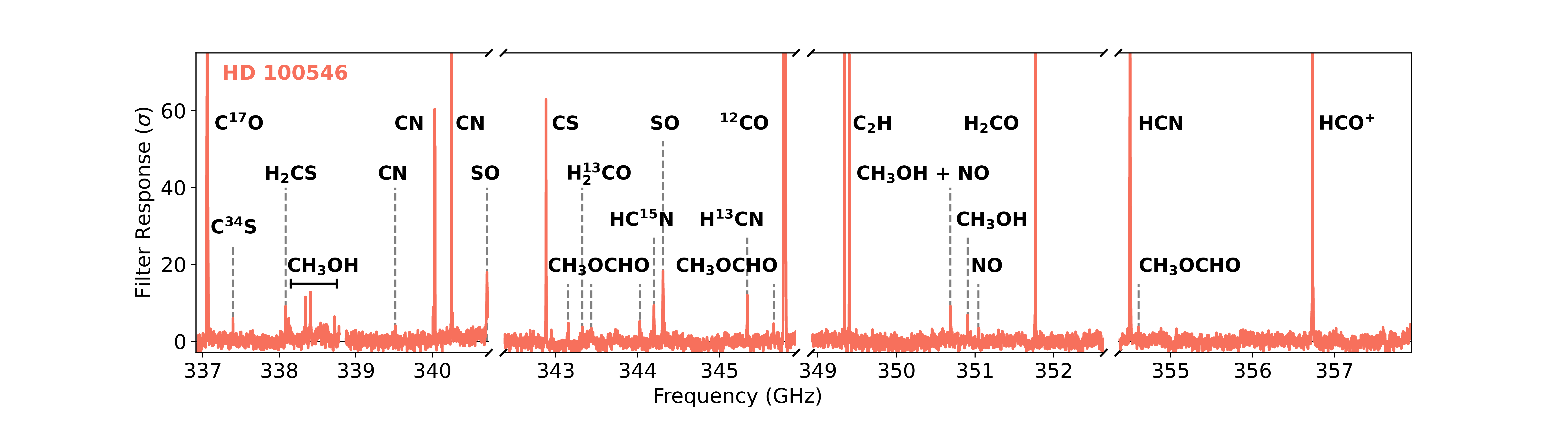}
    \caption{Matched filter response for the HD 100546 disk generated using Keplerian models with an outer radius of 300~au. Note the lines reaching the top of the y-axis have responses $>$70$\sigma$.}
    \label{fig:1}
\end{figure*}

\subsection{Data reduction}
HD~100546 was observed with ALMA in program 2021.1.00738.S (PI. A. S. Booth). These observations were taken in configuration C4 in Band 7 with baselines ranging from $\approx$15 to $\approx$1300~m.
These $\approx$0\farcs3 data are sufficient to distinguish between gas in the cavity and associated to the two dust rings but cannot resolve the gas within the cavity. 
The data consist of two spectral settings with four spectral windows each at a spectral resolution of 976.6~kHz (0.84~km~s$^{-1}$ at 350~GHz) and a bandwidth of 1.875~GHz. These spectral windows are centered at 338.790824, 340.732413, 348.916936 and 350.775389~GHz for setting A and, 344.240980, 3459.40999, 354.367095 and 356.067114~GHz for setting B. 
Further details on the spectral set-up and individual execution blocks are provided in the Appendix in Table~\ref{tab:A3}. 
This program also targeted the IRS~48 disk and these data will be presented in \citealt{Booth2023_irs48}.
The data were calibrated via the standard ALMA pipeline using \texttt{CASA} version 6.2.1 \citep{2007ASPC..376..127M}. These individually calibrated executions were aligned to a common phase centre using the \texttt{fixvis} and \texttt{fixplanets} tasks. Self-calibration was performed on the continuum data after flagging the strong lines. Both phase (6 rounds of decreasing solint intervals) and amplitude (1 round) calibrations were applied until the signal-to-noise of the continuum data plateaued. For HD~100546 this resulted in a peak continuum signal-to-noise increase from $\approx$525 to $\approx$5450. Self-calibration was necessary to detect the weak outer ring of millimetre emission in the HD~100546 disk. These solutions were then applied to the line data. The continuum was subtracted using \texttt{uvcontsub} with a fit order of 1 and excluding channels of bright lines and atmospheric absorption features that were identified via visual inspection of the data with the \texttt{plotms} task. 

\subsection{Line identification}

For an initial line identification in the data, we applied matched filtering \citep{Loomis2018}. This technique utilises the predictable kinematic pattern of Keplerian line emission to detect weak line emission whereby a filter, e.g., Keplerian model or strong line detection in the same disk (a so called image filter), is cross-correlated with the visibility data. Figure~\ref{fig:1} shows the resulting matched filter response using a Keplerian model with an outer radius of 300 au, which illustrates the frequency coverage of our observations and the line-rich nature of the data. We take a line detection to be any peak above the 4~$\sigma$ level.
As the impulse response depends on how accurately the filter represents the underlying emission structure, different filters, e.g., Keplerian models of different sizes and image filters of different strong lines, will yield a different signal-to-noise for a detected line \citep{2020ApJ...893..101L}. This is investigated further for the detection of weaker lines in Section 2.5.

\subsection{Imaging}

The data were imaged in CASA using \texttt{tCLEAN} with the multiscale deconvolver. Individual lines were cleaned with Keplerian masks down to 4$\times$ the noise level of the dirty image. The Keplerian masks were constructed using the properties for the HD~100546 disk as listed in Table~\ref{tab:A1}. These masks were generated with the code from \citet{rich_teague_2020_4321137}. A channel width of 976.6~kHz for each of the eight spectral windows results in a range of velocity resolutions from 0.82 to 0.87~km~s$^{-1}$ over the 337 to 356~GHz range of the observations. Therefore, for consistency, all of the spectral windows were imaged using a velocity resolution of 0.9~km~s$^{-1}$. For the strong lines, a briggs robust of +0.5 was used and for the weaker lines +0.5 or +2.0 was used depending on the emission morphology, e.g., compact or extended, respectively. The resulting beam sizes and rms noise for each line imaged are listed in Table~\ref{tab:hd100546_images} in the Appendix. 


\subsection{Weak-line detections}

To search for and confirm the detections of weak lines in the data we use the visibility and image plane techniques: matched filtering and spectral stacking with \textit{GoFish} (\citealt[][]{Loomis2018} and \citealt{GoFish}, respectively). For the matched filtering, as well as the Keplerian model result shown in Figure~\ref{fig:1}, we also use specific strong lines as filters to look for weaker lines of molecules that are either isotopologues of the strong line or are expected to share a common emission morphology. We also imaged the potentially weak line detections and used the spectral stacking tool \textit{GoFish} to create average spectra over different radial regions of the disks. 

\subsection{Column density calculations}

We estimate column densities following the methods outlined in \citet{2018ApJ...859..131L}. 
This assumes that the gas is in local thermodynamic equilibrium and includes an optical depth correction factor where we assume a line width of 1~km~s$^{-1}$.
Since we only have one detected transition for most molecules we do this over a set of fixed rotational temperatures. For the molecules where multiple transitions, e.g., \ce{CH_3OH}, are detected we pick one representative transition. Future work will focus specifically on constraining the excitation conditions of these molecules individually. In the case of CN, \ce{C_2H} and \ce{NO} the chosen lines are the strongest of the $N=3-2$, $N=4-3$ and $J=7/2-5/2$ hyper-fine groups, respectively. 
For \ce{SO_2} the $J=6_{(4,2)}-6_{(3,3)}$ is the strongest line detected and for SO the strongest transition is the $J=7_8-6_7$.
For \ce{CH_3OH} we pick the $J=7_{0}-6_{0}$ transition and for \ce{CH_3OCHO} and \ce{CH_3OCH_3} we use the $J=31-30$ and $J=19-18$ transitions which are both blends of multiple transitions. 

\begin{figure*}
    \centering
    \includegraphics[trim={0cm 0cm 0cm 0cm}, clip,width=\hsize]{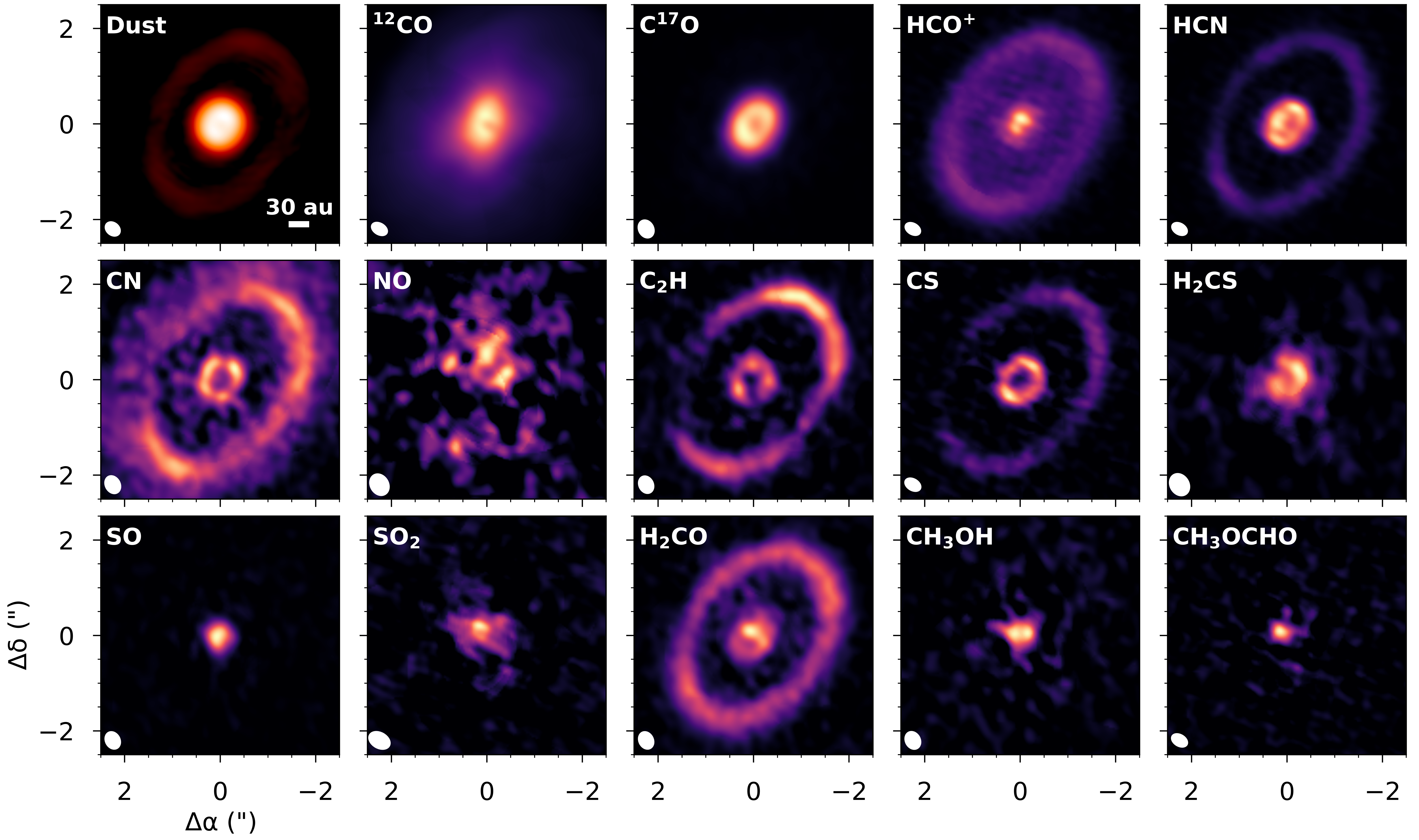}
    \caption{Integrated intensity maps of the 0.9~mm dust continuum emission and molecular line emission from the HD~100546 disk. The continuum map is shown on a log color scale to highlight the weak outer ring of millimetre emission. The beam is shown in the left-hand corner of each panel.}
    \label{fig:mom01}
\end{figure*}

\section{Results}

\subsection{Molecule detection summary}

In total, we have detected 19 different molecules in the HD~100546 disk. This includes the first detections of the rare isotopologues \ce{H_2^{13}CO} and \ce{^{34}SO} in protoplanetary disks. A range of nitrogen- (\ce{HCN}, \ce{CN}, \ce{NO}) and sulphur- (\ce{SO}, \ce{SO_2}, \ce{CS}, \ce{H_2CS}) bearing species are detected including some of the isotopologues along with \ce{^{12}CO}, \ce{C^{17}O}, \ce{HCO^+} and \ce{H_2CO}. Multiple \ce{CH_3OH} transitions are detected and we also robustly detect \ce{CH_3OCHO} (methyl formate). The larger organic molecules commonly detected in other disks, e.g., \ce{c-C_3H_2}, \ce{HC_3N}, \ce{CH_3CN} \citep{2021ApJS..257....9I}, remain undetected in our observations and this is primarily due to a combination of line sensitivity and the intrinsic properties of the lines covered.

\subsection{Weak-line detections / non-detections}

The results of the search for weak-line detections in the HD~100546 disk are summarised below and the associated figures are shown in Figure~\ref{fig:weaklines} in the Appendix. 

\begin{itemize}
    \item Using CS as a filter we detect \ce{C^{34}S} with a 9~$\sigma$ significance compared to the 6~$\sigma$ with the smooth 300~au Keplerian filter. 
    \item The \ce{H_2CO} filter detects \ce{H_{2}^{13}CO} at the 7~$\sigma$ level and the GoFish profile shows weak emission in the outer disk on 100-300~au scales.
    \item The \ce{HCO^+} filter does not detect \ce{HC^{18}O^+}.
    \item To search for NO we compared the results of a few different image filters and found \ce{H_2CO} to have the highest response and detect three NO lines (two of which are blended) at 5 and 4~$\sigma$, respectively. There are additional NO transitions covered in these data at 350.689~GHz but they are blended with a \ce{CH_3OH} line (as was also seen for the IRS~48 disk by \citealt{2022A&A...659A..29B, 2023arXiv230300768L}).
    \item With the SO filter we tentatively detect the \ce{^{34}SO} $J=9_8-8_7$ transition at the 4~$\sigma$ level. In the channel maps with a Briggs robust of +0.5, compact \ce{^{34}SO} emission is detected at the 3~$\sigma$ level across 5 consecutive channels which correspond to the same velocity range in which the SO is brightest. The other \ce{^{34}SO} line we cover ($J=8_8-7_7$) is not detected but the noise is higher in the region of this line than for the detected line (see Table~\ref{tab:hd100546_images}).
    We also report the non-detection of two \ce{OCS} lines. 
     \item Due to the strong detection of \ce{SO} in HD~100546 we could expect to also detect \ce{SO_2} since these molecules share a common gas-phase formation pathway via OH \citep[e.g.,][]{2018A&A...617A..28S}. No \ce{SO_2} lines are robustly detected with matched filtering using neither a Keplerian model nor an image filter. Since matched filtering works best for detecting weak emission that is spatially extended and spectrally resolved, if the \ce{SO_2} is compact like the \ce{SO}, it may not be detected with this technique. Indeed, in the image plane, 3 \ce{SO_2} lines with upper energy levels of $\approx$60-70~K are detected with GoFish. The strongest for which is the $J=6_{(4,2)}-6_{(3,3)}$ line.
    \item With the strongest CN, HCN and \ce{C_2H} lines as filters we detect weaker hyper-fine components of these transitions and also report non-detections of \ce{HC_3N}, \ce{c-C_3H_2} and \ce{CH_3CN}.
 \end{itemize}

\subsection{Integrated intensity maps}

Figure~\ref{fig:mom01} presents the integrated intensity maps of the representative transitions of each molecule (not all isotopologues) detected in the HD~100546 disk and the 0.9~mm continuum emission. The integrated intensity maps of the SO, CS and \ce{H_2CO} isotopologues are shown in Figure~\ref{fig:weaklines} in the Appendix and the HCN isotopologues will be the focus of future work. These line maps were generated using the Keplerian masks from the CLEAN-ing with no clipping thresholds. There are clear ring structures with radii of $\approx$200~au in the \ce{HCO^+}, \ce{CN}, \ce{HCN}, \ce{C_2H}, \ce{CS} and \ce{H_2CO} emission maps whereas the \ce{^{12}CO} shows no clear signs of sub-structure on these scales. The bulk of the \ce{C^{17}O} emission is compact with evidence of a central cavity at $<$20~au, also seen in high resolution continuum data \citep[e.g.,][]{2019ApJ...871...48P}, and the emission drops off steeply beyond $\approx$100~au. In comparison, the \ce{HCO^+} emission is centrally peaked. In the outer disk, the molecular rings appear to be approximately co-spatial with the weak millimetre dust ring. In some of the rings, there is an azimuthal variation in brightness (e.g. CN and CS) most strongly seen in the \ce{C_2H}. The detection of NO is weak but it also appears to be ringed. The \ce{SO} and \ce{CH_3OH} show compact emission as reported in \citet{2021NatAs...5..684B, 2023A&A...669A..53B} and the new detections of \ce{SO_2} and \ce{CH_3OCHO} are compact as well. 

\begin{figure*}
    \centering
    \includegraphics[width=0.93\hsize]{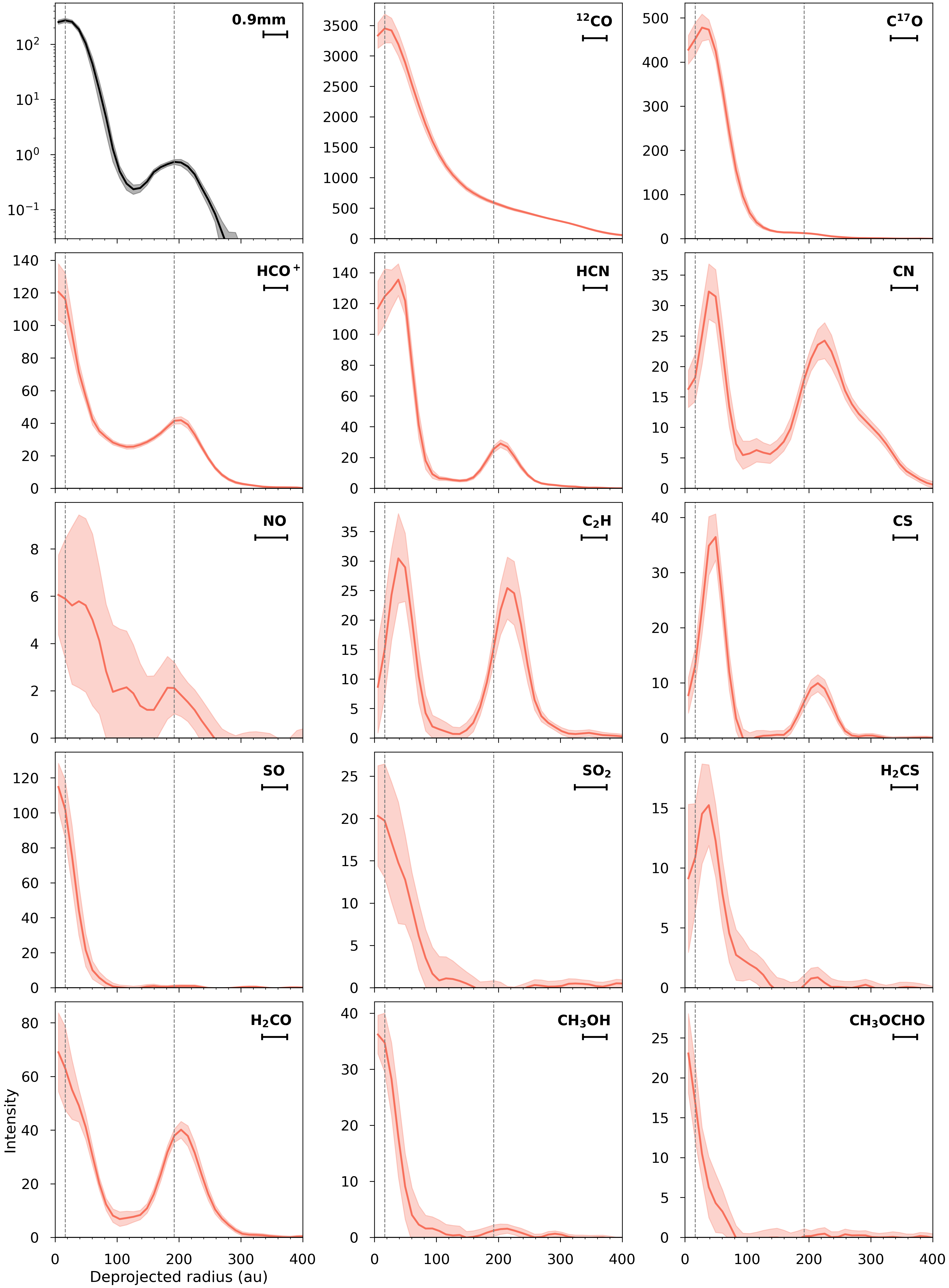}
    \caption{Azimuthally averaged radial emission profiles for the HD~100546 disk generated from the maps presented in Figure~\ref{fig:mom01}. The dashed vertical lines mark the location of peak intensity for the dust rings. The units of the y-axis are mJy~beam$^{-1}$~km~$\mathrm{s^{-1}}$ for the molecular lines and mJy~beam$^{-1}$ for the continuum which is shown on a logscale. The horizontal bars show the semi-major axis of the beam for each tracer.}
    \label{fig:HD100546_radial_profiles}
\end{figure*}

\subsection{Sub-structures in the HD~100546 disk}

Many of the integrated intensity maps of the molecular lines detected in the HD~100546 disk show clear radial and azimuthal substructures. In particular, the outer molecular rings are at a similar location to the outer dust ring, which to date, has not typically been seen in disk observations \citep[e.g.,][]{2021ApJS..257....3L}. In this section, we describe these different features and the possible physical and/or chemical processes responsible for generating these different sub-structures will be discussed in Section~4. 

Figure~\ref{fig:HD100546_radial_profiles} shows the azimuthally averaged radial intensity profiles for the lines shown in Figure~\ref{fig:mom01}. The gas disk traced by \ce{^{12}CO} is 600~au in size which is $\approx$2$\times$ the radial extent of the millimetre continuum emission. The CO isotopologues show a cavity in the inner disk and while there is no apparent gap in the outer disk traced in \ce{^{12}CO} there is a shelf of \ce{C^{17}O} emission across the location of the outer dust ring.
Interestingly, the \ce{HCO^+} emission is centrally peaked and also shows a clear gap that follows the dust rings. 
The array of molecular rings at $\approx$200 to 240~au clearly align with the location of the outer dust ring but there are some shifts between different molecules, e.g., the HCN and CN peaks are offset by approximately half a beam from each other.  The line flux within the outer dust gap at $\approx$110~au also varies between molecules where \ce{HCO+} is the least reduced and the deepest gaps are found for \ce{C_2H} and \ce{CS}. There is also variation in the relative brightness of the two rings in different molecules. For example, the \ce{C_2H} and \ce{CN} rings in the inner and outer disk are similarly bright, whereas the \ce{HCN}, \ce{HCO^+}, \ce{CS}, \ce{NO} and \ce{H_2CO} rings are $\approx$2-5$\times$ fainter in the outer disk. In the inner disk, there is a clear anti-correlation between the SO (and \ce{SO_2}) and the CS (and \ce{H_2CS}). Weak rings of emission are present in the radial profiles for SO (also shown more clearly in \citealt{2023A&A...669A..53B}) and \ce{CH_3OH} (tentatively detected in \citealt{2021NatAs...5..684B}). To check more robustly for weak extended emission from other molecules we use GoFish \citep{GoFish} to create average spectra stacked within an annulus, of 180-250~au, which encompasses the emission  detected in the outer ring. With this analysis, the outer rings of \ce{SO} and \ce{CH_3OH} are confirmed along with outer rings of \ce{H_2CS} and \ce{^{34}CS}. Rings of \ce{SO_2} and \ce{CH_3OCHO} both go undetected. These six spectra are shown in Figure~\ref{fig:gofish} in the Appendix. 

\begin{figure}
    \centering
    \includegraphics[width=\hsize]{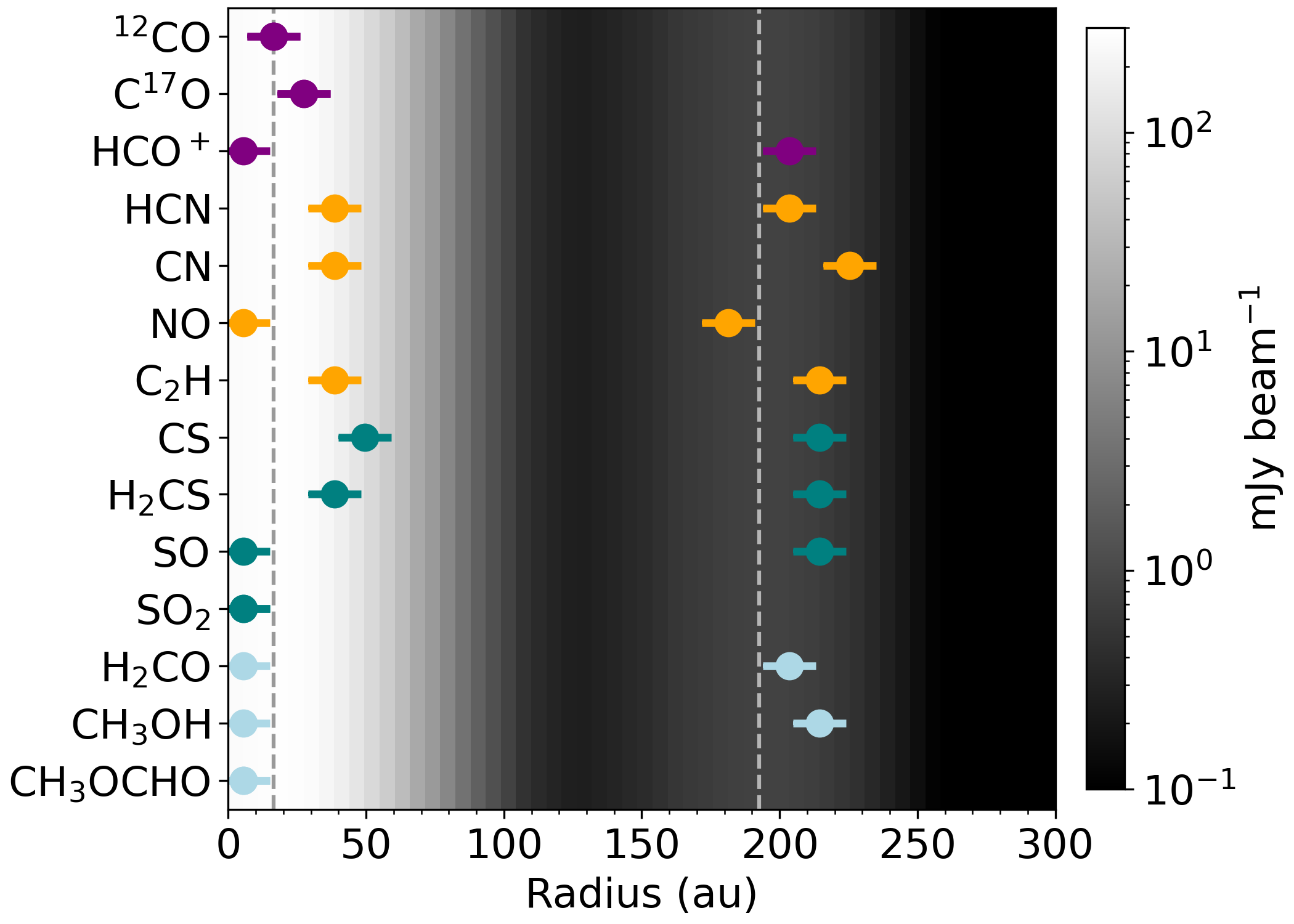}
    \caption{Colour map of the azimuthally averaged dust continuum emission for the HD~100546 disk. The coloured markers show the peak location of the different molecular rings where the width of the horizontal lines show 0.5$\times$ the semi-major axis of the beam. Different colors show different elemental/molecular families. The dashed vertical lines mark the location of peak intensity for the dust rings.}
    \label{fig:radial_scatter}
\end{figure}

To examine the relative locations of these rings in more detail we plot the location of the ring peaks for each molecule in Figure~\ref{fig:radial_scatter} overlaid on a map of the azimuthally averaged radial profile of the continuum emission. There is a clear correlation between the positions of the peaks of the dust rings and the peaks of the molecular rings. Note this alignment of molecular rings with the dust is similar to what was seen in the PDS~70 disk \citep{2021AJ....162...99F} and in the HD~163296 disk for the dust ring at 101~au \citep{2021ApJS..257....3L}. For the inner disk, all the molecules that contain oxygen (except \ce{C^{17}O}) peak within the central dust cavity and other molecules peak just outside center of the first dust ring. In the outer disk, the molecules all align with the second dust ring at 200~au but there is a gradient, e.g., CN and \ce{C_2H} reach a brightness maximum farther out radially than the HCN and \ce{HCO^+}. 

\begin{figure*}
    \includegraphics[width=0.9\hsize]{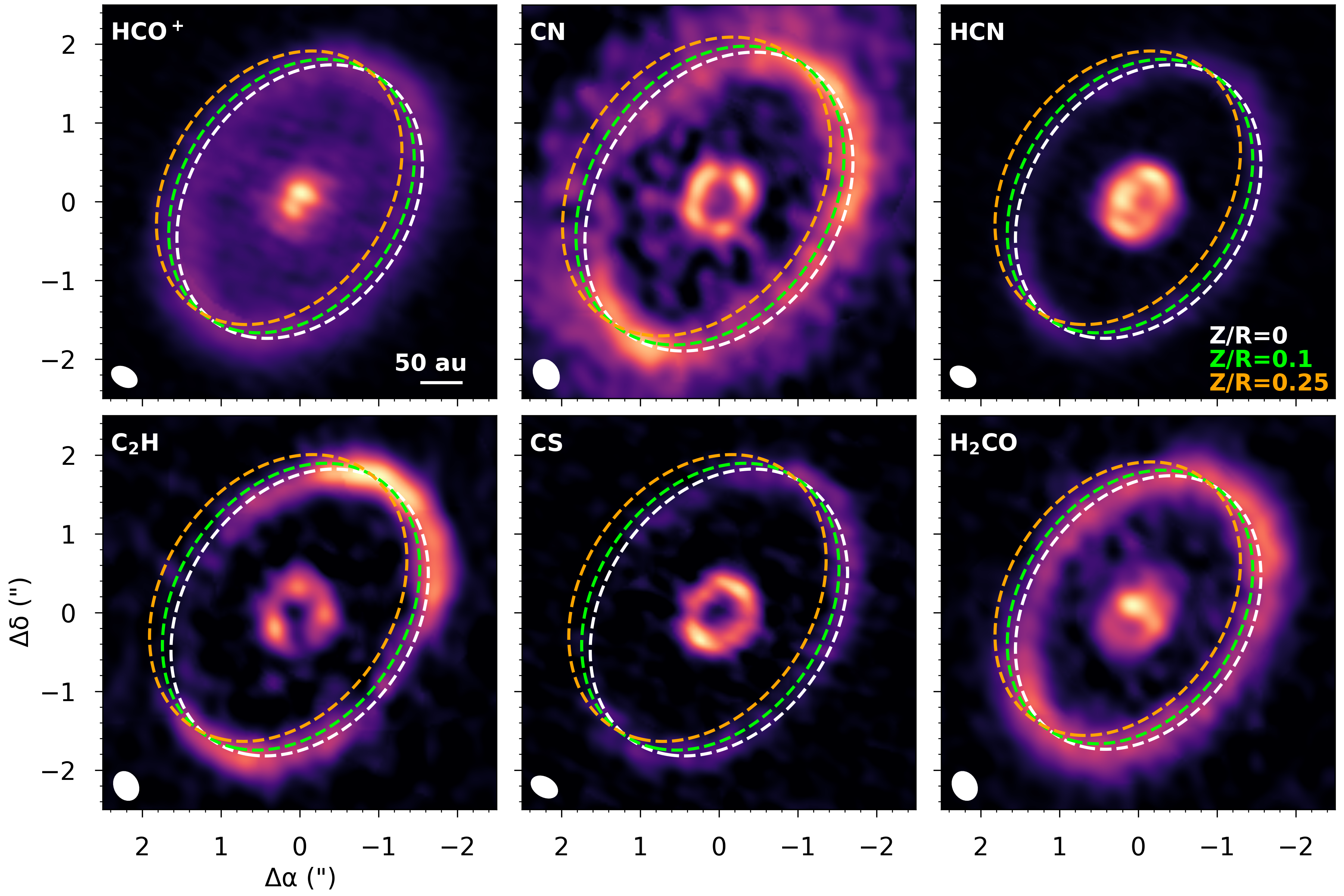}
    \caption{Integrated intensity maps of molecular line emission from the HD~100546 disk with three ellipses overlaid showing the expected morphology of a ring of emission emitting from different heights (Z/R) in the disk. A Z/R=0 is effectively flat, i.e., emitting from the midplane, and Z/R=0.25 is the most elevated and this is the measured surface for the \ce{^{12}CO} at $\approx$200~au from both \citet{2022ApJ...932..114L} and  \citet{2023A&A...669A.158S}.}
    \label{fig:ellipses}
\end{figure*}

To explore whether or not these differences in peak radial locations are due to a different emitting height of these molecules in the disk we overlay ellipses on the integrated intensity maps of \ce{HCO^+}, HCN, CN, \ce{C_2H}, CS and \ce{H_2CO} (see Figure~\ref{fig:ellipses}). The ellipses are tracing the Z/R (geometric height from the midplane divided by radius) of 0 (the midplane), 0.1 and 0.25 surfaces at the radius where each of the species peaks in emission. Although we do not have the kinematic resolution to directly measure the emitting heights of these different molecules from the channel maps (e.g., \citealt{2023A&A...669A.126P}) we can gain insight from the geometry of the molecular rings compared with the different Z/R ellipses. 
The molecular rings in the outer disk appear to be arising from a layer in the disk close to the midplane and, given the spatial resolution of our observations, we can constrain this to be from a layer with Z/R$<$0.1. This is in contrast with the measurements of the \ce{^{12}CO} emitting height where both \citet{2022ApJ...932..114L} and \citet{2023A&A...669A.158S} derive a Z/R of $\approx$0.25 at 200~au. Therefore, different emitting heights are likely not responsible for the different ring peak locations. 
For the inner disk, $<$50~au, we do not have the spatial resolution to rule out that these inner rings are arising from different layers in the disk. 

\begin{figure*}
    \includegraphics[width=0.9\hsize]{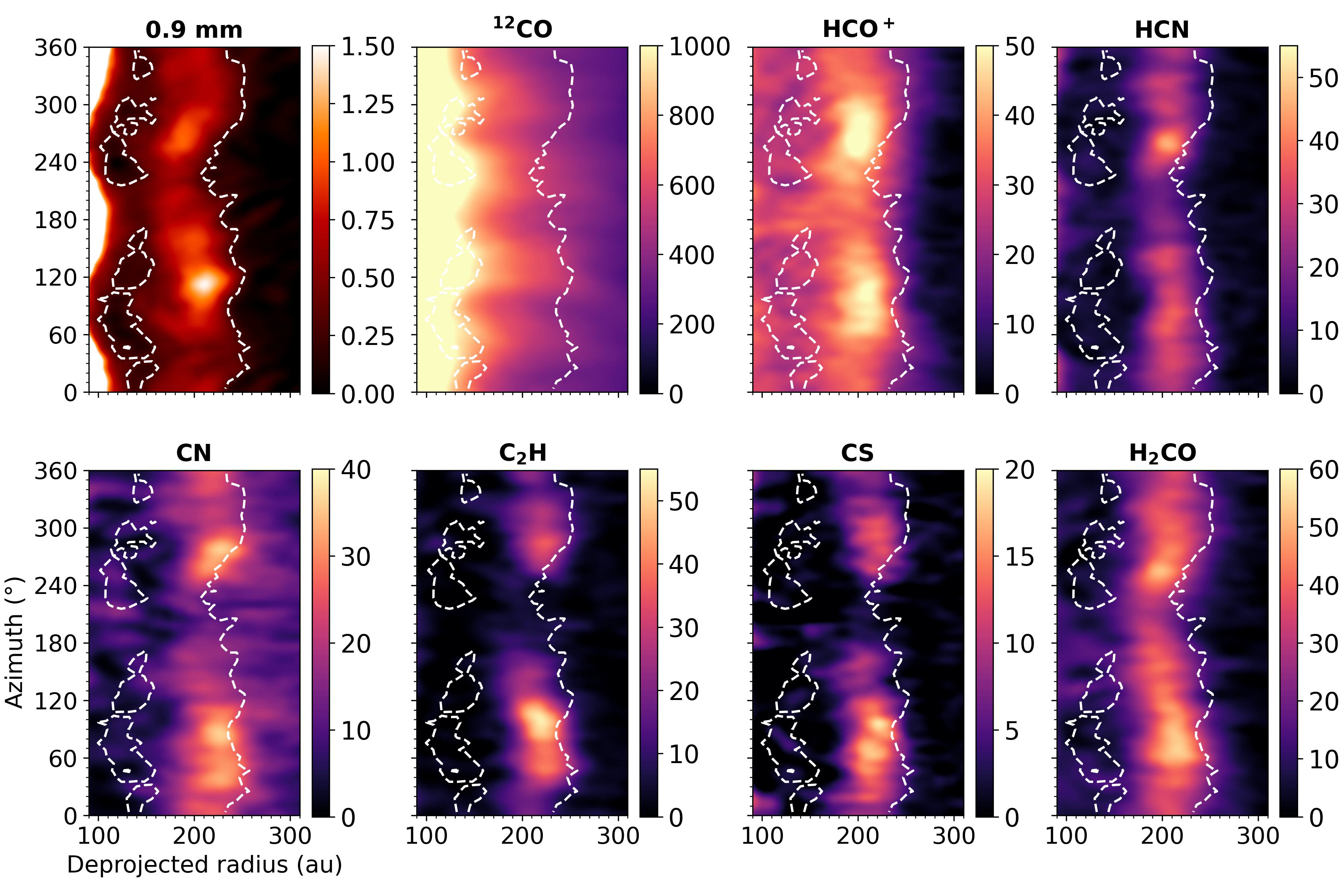}
    \caption{Polar de-projection of the HD~100546 integrated intensity maps which highlight the azimuthal asymmetry in the dust and molecular rings at $\approx$200~au. 
    The dashed contour traces the 5$\sigma$ level of the dust continuum emission. The units of the color bar are mJy~beam$^{-1}$~km~$\mathrm{s^{-1}}$ for the molecular lines and mJy~beam$^{-1}$ for the continuum.}
    \label{fig:hd100546_azimuth}
\end{figure*}

There are also azimuthal asymmetries in the brightness of the outer molecular rings in HD~100546. This asymmetry is most clearly seen in \ce{C_2H}, CN and CS. In Figure~\ref{fig:hd100546_azimuth} we show polar deprojections of the integrated intensity maps for the 0.9~mm continuum, \ce{^{12}CO}, \ce{HCO^+}, \ce{CN}, HCN, \ce{C_2H}, CS and \ce{H_2CO}. From this visualisation, it is clear that the molecules peak in similar azimuthal regions of the disk and this also follows an underlying asymmetry in the dust emission (as reported in \citealt{2021A&A...651A..90F}). There is a tentative eccentricity in the molecular line emission, seen most clearly in the \ce{H_2CO}, which is also seen in the dust by \citealt{2021A&A...651A..90F}. 

\begin{figure*}
    \includegraphics[width=0.9\hsize]{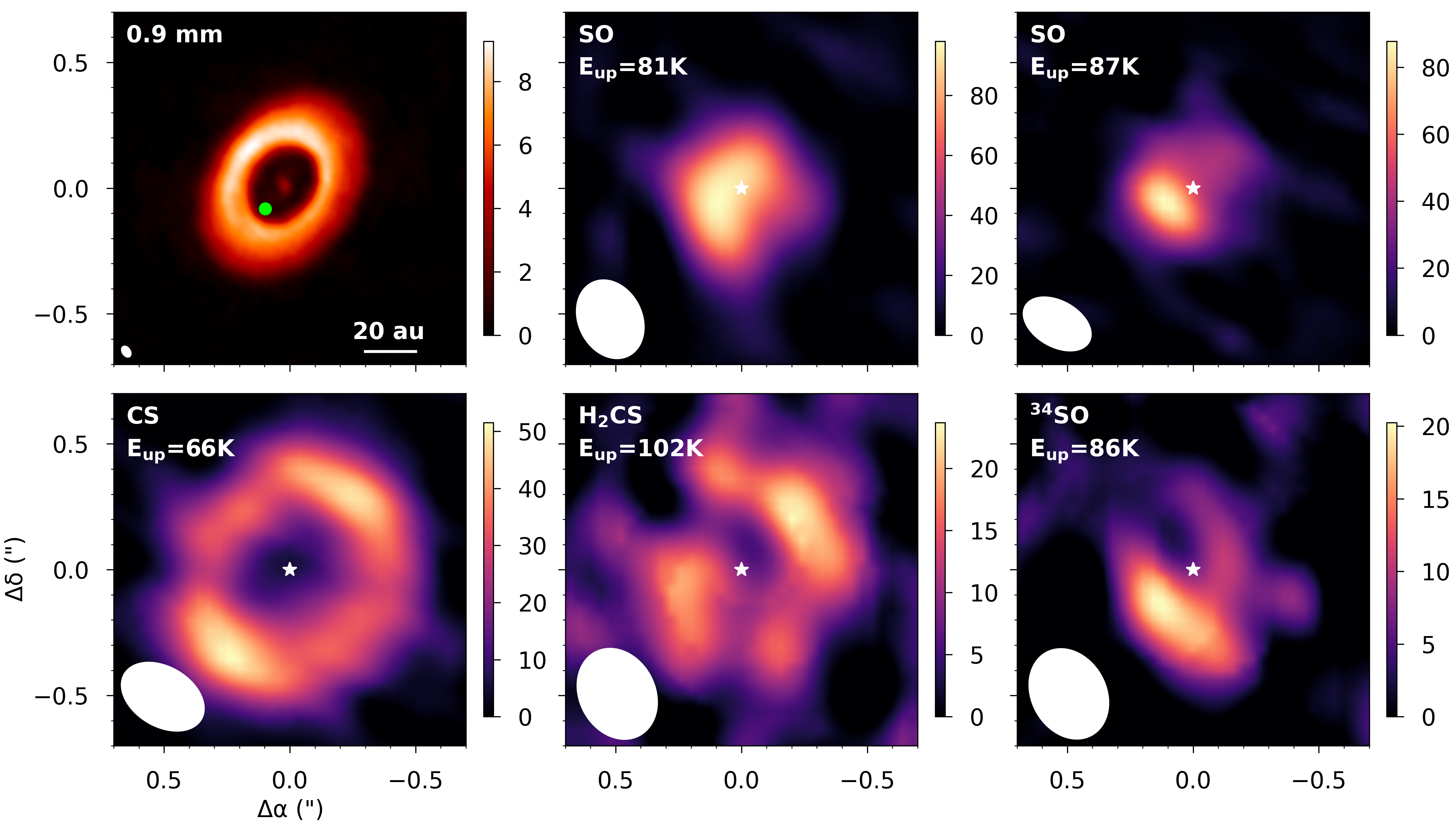}
    \caption{A gallery of sulphur-bearing molecules detected in the HD~100546 disk alongside the 0.9~mm dust emission from \citet{2019ApJ...871...48P}.  The location of the putative planet from \citet{2015ApJ...814L..27C} and the excess ro-vibrational CO emission attributed to a forming planet by \citet{2019ApJ...883...37B} from 2017 are highlighted with a green circle.
    This is the same region of the disk where the SO emission is peaking. The units of the color bar are mJy~beam$^{-1}$~km~$\mathrm{s^{-1}}$ for the molecular lines and mJy~beam$^{-1}$ for the continuum.}
    \label{fig:hd100546_sulphur}
\end{figure*}

\citet{2023A&A...669A..53B} reported asymmetric SO within the central dust cavity of the HD~100546 disk. In this work, we present two additional SO lines along with the other sulphur species CS and \ce{H_2CS}. 
Figure~\ref{fig:hd100546_sulphur} shows the integrated intensity maps of the dust emission from \citealt{2019ApJ...871...48P} and the two new SO transitions detected in this work ($J=7_8-6_7$ and $J=8_8-7_7$) alongside the CS $J=7-6$, \ce{H_2CS} $J=10_{(1,10)}-9_{(1,9)}$ and, the detection of \ce{^{34}SO} $J=9_8-8_7$. From this image gallery it is clear that the SO is sitting inside a broadly symmetric ring of CS and the SO asymmetry is most prominent in the warmest line we have detected (this is also the line with the smallest restoring beam). In addition, the \ce{^{34}SO} is only detected in the south of the cavity where the main SO isotopologue is the brightest. 

\subsection{Column densities}

\begin{figure*}
    \centering
    \includegraphics[width=0.93\hsize]{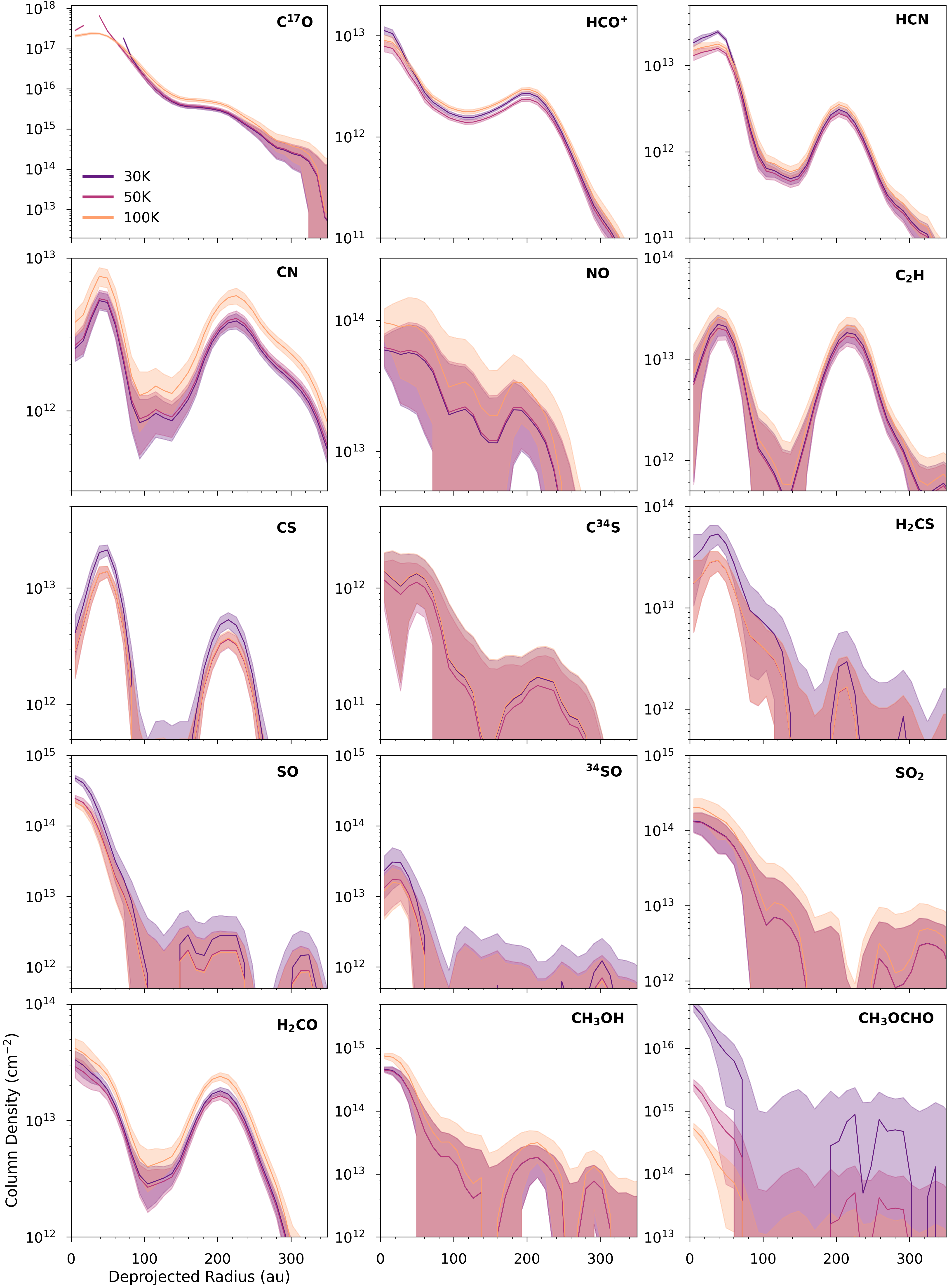}
    \caption{Radial column density profiles for the HD~100546 disk determined at a range of assumed excitation temperatures. The \ce{C^{17}O} profiles break where the optical depth exceeds $\approx$2.}
    \label{fig:hd100546_radial_columns}
\end{figure*}

We compute radial column density profiles for the molecules in the HD~100546 disk from the profiles presented in Figure~\ref{fig:HD100546_radial_profiles}.
We explore a range of excitation temperatures: 30, 50, and 100~K and the results are shown in Figure~\ref{fig:hd100546_radial_columns} and we summarise our findings here. 

\begin{itemize}
    \item \ce{C^{17}O} reaches a peak column density of the order of $10^{17}~\mathrm{cm^{-2}}$ but becomes optically thick with a temperature of 50~K within 40~au and approaching optically thick at 100~K ($\tau\approx0.6$). In Table~\ref{tab:ratios} we list the column density ratios of different molecules relative to CO at their peak locations in the inner disk (at 100~K) and peak in the outer disk (at 30~K and 50~K). The \ce{C^{16}O} column density is determined from the \ce{C^{17}O} column density assuming \ce{^{16}O}/\ce{^{18}O}=557 and \ce{^{18}O}/\ce{^{17}O}=3.6
    \citep{1999RPPh...62..143W}. For the inner disk these column density ratios are tentative due to potentially optically thick emission. 
    
    \item \ce{HCO^+} and HCN are both approaching optically thick in the inner 40~au of the disk with the peak value for $\tau$ ranging from 0.1 to 1.0 depending on the gas temperature, and have similar overall column densities. In comparison, the CN, NO and \ce{C_2H} are optically thin throughout the disk and NO has the highest column density of the three nitrogen-bearing molecules detected.
    
    \item For both CS and SO we can assess the optical depth of the main isotopologues with the detections of \ce{C^{34}S} and \ce{^{34}SO} where in the local ISM \ce{^{32}S}/\ce{^{34}S} is 22 \citep{1999RPPh...62..143W}.  In the inner disk where the CS and SO peak, the respective column densities of the \ce{^{34}S} isotopologues can be up to 3$\times$ higher than the main isotopologue once correcting for the isotope ratio. This indicates optically thick emission from the main isotopologues in the inner disk or an enhanced \ce{^{34}S} isotope abundance. 
    In the outer ring at 220~au N(CS)$\approx$N(\ce{C^{34}S})$\times$22 which indicates optically thin CS emission if the S isotope ratio is indeed the local interstellar value. The \ce{CS}/\ce{H_2CS} ratio is $\approx2\pm$2 in this outer ring. For the inner disk we use the optically thin \ce{C^{34}S} and with \ce{C^{34}S}$\times$22 we find a  \ce{CS}/\ce{H_2CS} ratio of 1$\pm$0.5 in the inner ring. For \ce{SO}/\ce{SO_2} using the \ce{^{34}SO} results in a column density ratio of $\approx$2.0$\pm$1.0 in the inner disk. 
    
    \item For the larger organics \ce{H_2CO}, \ce{CH_3OH} and \ce{CH_3OCHO} in the inner disk the lines appear to be optically thin but may be beam diluted. The peak column density ratios in the inner disk for \ce{CH_3OH}/\ce{H_2CO} is $18\pm4$ and for \ce{CH_3OCHO}/\ce{CH_3OH} is 0.7$\pm$0.2 at 100~K for the inner disk. In the outer ring \ce{CH_3OH}/\ce{H_2CO} is $1.1\pm0.6$ at both 30 and 50~K.
\end{itemize}

\begin{table*}[]
    \centering
        \caption{Column density ratios of different molecules (X) relative to CO at the central peak or inner ring and, outer ring in the HD~100546 disk. Errors on radial locations are 0.5$\times$~the major axis of the beam.}
    \begin{tabular}{c  c  c  c c c } \hline \hline
      Molecule & Inner Peak & N(X)/N(CO) & Outer Ring Peak & N(X)/N(CO)  & N(X)/N(CO) \\
        & (au) &  $T_{ex}$=100~K & (au) &  $T_{ex}$=30~K  & $T_{ex}$=50~K   \\
    \hline
       \ce{HCO^+}   & 5.5$\pm$19 & $2.2\pm0.3\times10^{-8}$ & 203.5$\pm$19 & $4.8\pm0.5\times10^{-7}$ & $3.9\pm0.4\times10^{-7}$ \\
       \ce{HCN}     & 38.5$\pm$19 & $3.8\pm0.3\times10^{-8}$ & 214.5$\pm$19 & $6.0\pm0.8\times10^{-7}$ & $5.1\pm0.7\times10^{-7}$ \\
       \ce{CN}      & 38.5$\pm$21 &$1.6\pm0.2\times10^{-8}$ & 225.0$\pm$21 & $1.1\pm0.2\times10^{-6}$ & $1.1\pm0.2\times10^{-6}$ \\
       \ce{NO}      & 5.5$\pm$26 & $2.3\pm0.7\times10^{-7}$& 181.0$\pm$26 & $3.0\pm2.0\times10^{-6}$ & $3.0\pm2.0\times10^{-6}$ \\
       \ce{C_2H}    & 38.5$\pm$20& $6.0\pm1.0\times10^{-8}$& 214.5$\pm$20 & $5.0\pm1.0\times10^{-6}$ & $5.0\pm1.0\times10^{-6}$ \\
       \ce{CS}      & 49.5$\pm$19 & $3.4\pm0.4\times10^{-8}$ & 214.5$\pm$19  & $1.3\pm0.3\times10^{-6}$ & $9.0\pm2.0\times10^{-7}$ \\
       \ce{SO}      & 5.5$\pm$20& $5.2\pm0.7\times10^{-7}$ & 214.5$\pm$20 & $3.0\pm3.0\times10^{-7}$ & $6.0\pm5.0\times10^{-7}$ \\
       \ce{H_2CS}   & 38.5$\pm$21& $6.0\pm1.0\times10^{-8}$ & 214.5$\pm$21 & $6.0\pm5.0\times10^{-7}$ & $3.0\pm3.0\times10^{-7}$ \\
       \ce{H_2CO}   & 5.5$\pm$20 & $1.0\pm0.2\times10^{-7}$& 203.5$\pm$20 & $3.2\pm0.4\times10^{-6}$ & $3.2\pm0.4\times10^{-6}$ \\
       \ce{CH_3OH}  & 5.5$\pm$19 & $1.8\pm0.3\times10^{-6}$& 214.5$\pm$19 & $4.0\pm2.0\times10^{-6}$ & $4.0\pm2.0\times10^{-6}$ \\
       \hline \hline
    \end{tabular}
    \label{tab:ratios}
\end{table*}

\section{Discussion} \label{sec:discussion}

In this section we discuss the potential physical and/or chemical origins of the molecular sub-structures detected in the HD~100546 disk and 
compare this disk to other chemically well-characterized Herbig Ae disks. 

\subsection{The origin of the ringed molecular sub-structures in HD~100546}

The HD~100546 disk shows clear rings in all of the molecules detected aside from CO and the locations of these rings all correlate with the radii of the mm-dust rings. This is the first time such a clear connection between dust and molecular rings has been in the outer regions of a disk.   
The most straightforward explanation for this coincidence would be the presence of a $\approx$100~au wide and deep gas gap that has shaped the larger millimetre-sized dust. The hydrodynamical models of planet-disk interactions presented in \citet{2021A&A...651A..90F} that reproduce the mm-dust emission estimate approximately an order of magnitude drop in gas column density within the outer dust gap.  Although there is a shelf of \ce{C^{17}O} emission a
clear gap is column density not immediately apparent. This is in contrast to other systems with 10's of au wide dust gaps where corresponding CO gas gaps have been detected \citep[e.g., AS~209 and HD~169142;][]{2019ApJ...871..107F, 2022MNRAS.517.5942G}. 

Molecular rings can also arise in disks due to the disk chemistry rather than the disk's physical structure. The cyanides CN and HCN are expected to be ringed due to the interaction of the disk surface layer with the stellar UV radiation
\citep{2018A&A...609A..93C,2018A&A...615A..75V}. 
\ce{HCO^+} is also expected have a central depression where \ce{H_2O} is abundant in the gas-phase \citep{2021A&A...646A...3L} but interestingly, in HD~100546 we find centrally peaked \ce{HCO^+} emission from within the gas cavity. 
\citet{2022A&A...665A..45P} model the \ce{H_2O} lines detected/not-detected in the HD~100546 disk with \textit{Herschel} and suggest that the \ce{H_2O} in the central cavity has been photo-dissociated to form OH, therefore, this could allow for \ce{HCO^+} to be present. 
Finally, the hydrocarbon \ce{C_2H} is proposed to be abundant at the edge of the pebble disk and in the dust gaps \citep{2021ApJ...910....3B}.
These chemical rings are not all expected to arise at the same radial location as each other as seen in the data. 

The molecular rings we detect could also be shaped by radial variations in the disk gas-phase C/O. This would require an elevated C/O at the dust rings where the \ce{C_2H} is bright and a C/O$<$1 within the central dust cavity and in the outer gap. 
Cold \ce{H_2O} vapour has been detected within this outer gap region $\approx$40-150~au \citep{2021A&A...648A..24V, 2022A&A...665A..45P} and the detection of \ce{CH_3OH} within the central cavity would be consistent with this. This interpretation also fits with the central compact SO located inside a ring of CS where the switch between dominant gas-phase S-carrier occurs across the \ce{H_2O} snowline. This is because, if the SO and CS are forming via gas-phase reactions their relative abundances will reflect the underlying C/O \citep{2020A&A...638A.110F}.
It is not quite as simple as this though since SO, NO, \ce{H_2CO} and \ce{CH_3OH} are detected in the dust outer ring as well as the \ce{C_2H}, CS and CN. There are some radial offsets in the peak locations of these outer rings, as shown in Figures~\ref{fig:radial_scatter} and \ref{fig:hd100546_azimuth} and that along with the apparent flat nature of the emission can be investigated with disk chemical models. 
There is likely also a contribution of molecular column density to the outer ring from non-thermal desorption, particularly evident from the detection of \ce{CH_3OH} at $\approx$200~au and the cold \ce{H_2O} from \textit{Herschel}. The physical and chemical origin of these rings will be investigated further with grids of thermochemical models in Leemker et al. (2023, in prep) and the radial variation in \ce{CH_3OH}/\ce{H_2CO} and contribution from non-thermal desorption mechanisms will be investigated in Evans et al. (2023, in prep). 

\subsection{The origin of the asymmetric molecular sub-structures in HD~100546}

The molecular emission maps in Figure~\ref{fig:mom01} also show azimuthal asymmetries on large scales as highlighted most clearly in Figure~\ref{fig:hd100546_azimuth}. There is a drop in emission for all of the well-detected molecules in both the East and West of the disk that is also seen in the continuum emission.
The azimuthal asymmetries in the outer rings could be due to azimuthal variation in temperature, but the lack of clear signature in the optically thick \ce{^{12}CO} does not support this.  
Alternatively, the asymmetry could be the result of varying incident UV radiation due to self-shadowing from the mm-dust ring at 20~au, but this warrants further investigation \citep{2010A&A...519A.110P, 2017A&A...607A.114W, 2021MNRAS.505.4821Y, 2023NatAs...7..684K}.
\citet{2021A&A...651A..90F}, who have higher angular resolution continuum observations than are presented here, show that the outer dust ring is clumpy and eccentric. Our line observations appear to trace this sub-structure but $\approx$0.1" line data are needed to understand the connection between the gas and the large dust in the disk midplane. All and all, this could point to an azimuthal variation in molecular column density that is following the dust. 

On smaller, $<$1" scales, we recover the SO spatial asymmetry presented by \citet{Booth2023}. The SO (and \ce{^{34}SO}) are brightest in the south of the disk where excess CO ro-vibrational emission was detected with VLT/CRIRES and where a disk-feature was detected in scattered light (\citealt{2019ApJ...883...37B,2015ApJ...814L..27C}; see Figure~\ref{fig:hd100546_sulphur}). Additionally, the two new SO lines detected show the same asymmetric line profile as the two lines from \citet{Booth2023}. The lower signal-to-noise \ce{SO_2} emission does not show a clear asymmetry in the integrated intensity maps but the emission in the channel maps appears to be brightest in the south of the disk. Further, high-angular resolution, observations particularly of shock-tracing species such as SiO should help elucidate the nature of this emission and if it is indeed related to a forming giant planet \citep[as eluded to in HD~169142;][]{2023arXiv230613710L}.


\subsection{In context with other Herbig Ae disks}

Here we compare the HD~100546 disk to the HD~163296 and MWC~480 disks which were both observed in the ALMA Large Program MAPS \citep{2021ApJS..257....1O}. We also include some comparisons to the asymmetric transition disk HD~142527 \citep{2023A&A...675A.131T} and the ringed transition disk HD~169142 \citep{Booth2023} where applicable. There are also now some efforts to survey the chemistry in Herbig Ae disks, but we do not cover the same molecules/transitions as these works \citep{2022MNRAS.510.1148S, 2023arXiv230302167P}. We will therefore mainly focus on the above individual disks where column densities have been calculated and the observations are spatially resolved. 

\begin{figure*}
    \centering
\includegraphics[width=\hsize]{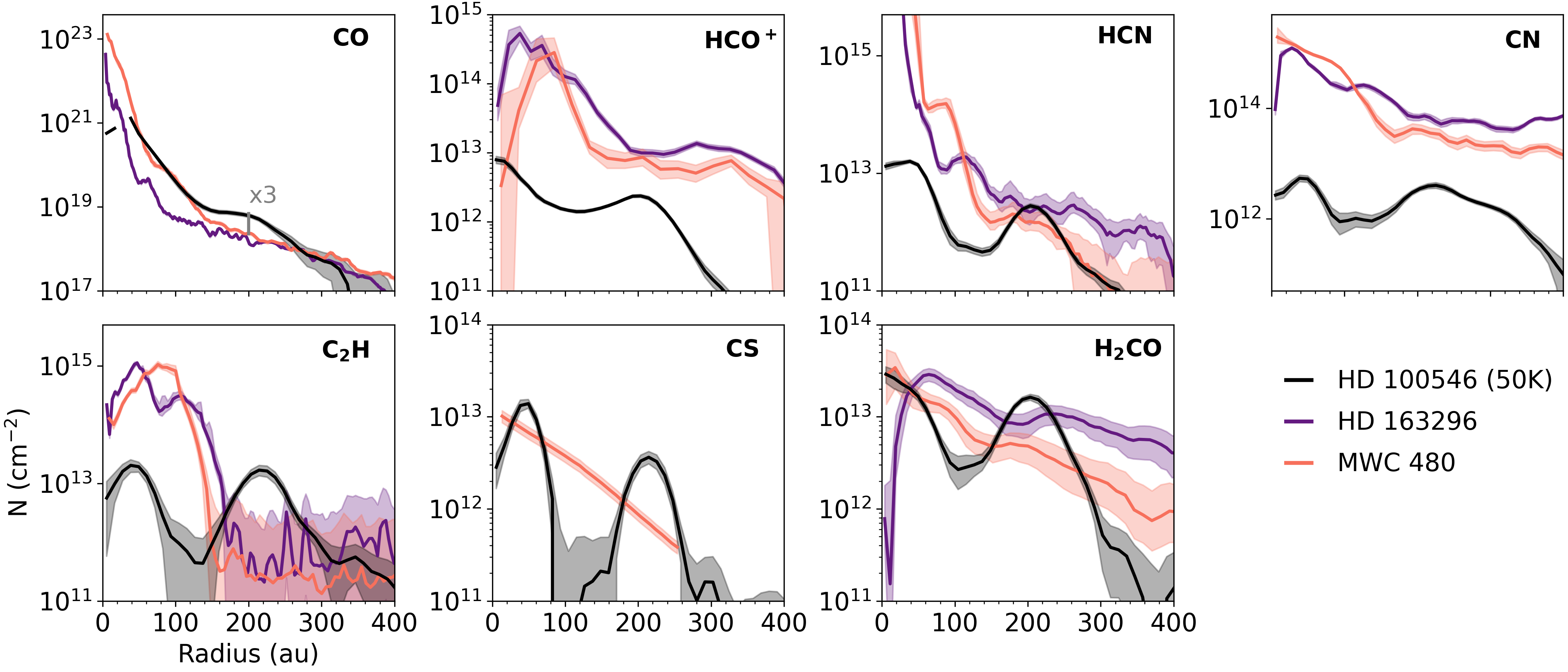}
    \caption{A comparison of the radial column density profiles in HD~100546 (at 50K) and the radial column densities of the two Herbig Ae disks - HD~163296 and MWC 480 - observed in the ALMA large program MAPS \citep{2021ApJS..257....1O, 2021ApJS..257....5Z, 2021ApJS..257...13A, 2021ApJS..257...12L, 2021ApJS..257....6G}.}
    \label{fig:comparison}
\end{figure*}

\subsubsection{Disk physical structure} 

The dust masses, gas masses and the gas disk sizes of the HD~100546, HD~163296 and MWC~480 disks are all similar \citep{2021ApJS..257...14S, 2021ApJS..257....5Z, 2021ApJS..257...17C, 2021ApJS..257....3L}. One notable difference in physical structure between these three disks is that the HD~100546 disk is warmer with no evidence for significant CO freeze-out \citep[see models from][]{2016A&A...592A..83K}.  In comparison, the HD~163296 and MWC~480 disks each have a midplane CO snowline at 65 and 100~au, respectively, where beyond this radius gas-phase CO is depleted by $\approx$10$\times$. This temperature difference may be due to the large central cavity in the HD~100546 disk resulting in a warmer midplane throughout the disk. Figure~\ref{fig:comparison} shows the CO column density in the HD~100546 disk determined from \ce{C^{17}O} derived under the assumption of ISM isotope ratios compared to that for HD~163296 and MWC~480 from \citet{2021ApJS..257....5Z}. In the outer disk, at the location of the HD~100546 second dust ring, the CO column density is $\approx3\times$ higher in HD~100546 compared to the other two disks at the same radii. This is likely due to less efficient CO depletion in this warmer disk \citep{2018A&A...618A.182B}. The actual CO column could be higher by an additional factor of $\approx3$ if the \ce{C^{17}O} abundance is underestimated due to isotope-selective photo-dissociation \citep{2016A&A...594A..85M}. So although these disks have a similar mass and size there is significantly more gas-phase CO available in the HD~100546 disk particularly at the location of the observed chemical rings in the outer disk. 

\subsubsection{Ionization} 

The primary tracer of ionization in disks in the warm molecular layer is the cation \ce{HCO+}. \ce{HCO+} forms via the reaction of CO with \ce{H_3^+} which forms via the ionization of \ce{H_2}. The main destruction routes of \ce{HCO^+} are dissociative recombination with an electron or a charge transfer reaction with \ce{H_2O} \citep[e.g.,][]{2021A&A...646A...3L}. The radial \ce{HCO^+} column density in the HD~100546 disk compared to HD~163296 and MWC~480 is shown in Figure~\ref{fig:comparison} where the data for the other two disks are taken from \citet{2021ApJS..257...13A}. Overall, the \ce{HCO^+} column density is generally lower in HD~100546 disk by a factor of $\approx$10. The ratio \ce{HCO^+}/\ce{CO} can be used as a proxy for the degree of ionization in the disk and \citet{2021ApJS..257...13A} found that for HD~163296 and MWC~480 this value ranges from $\approx10^{-5}$ to $\approx10^{-4}$ and $\approx10^{-6}$ to $\approx10^{-5}$ from 100 to 400~au for each of these two disks respectively which is broadly in agreement with X-ray ionization disk chemical models. In HD~100546, we find that at $\approx$200~au (where \ce{C^{17}O} and \ce{HCO^+} are optically thin) this is at least an order of magnitude lower - $3-5\times10^{-7}$ (see Table~2). This lower value is similar to the HD~142527 disk where N(\ce{HCO^+})/N(CO) are $\approx2\times10^{-7}$ \citep[][priv. comm.]{2023A&A...675A.131T}. The primary sources of ionisation in disks are stellar X-rays in the warm molecular layer and cosmic rays and short-lived radionuclides (SLRs) in the disk midplane. The X-ray luminosity of HD~100546 is a factor of 2-5$\times$ lower than that of HD~163296 and MWC~480 \citep[see Table 1 and][]{2019A&A...625A..66D}. There is currently no evidence to suggest differences in the cosmic ray ionisation rates or SLRs abundances between these three sources; hence, the difference in stellar X-ray properties could account for the different \ce{HCO^+} abundances but is likely not the whole story. In addition to a lower ionization rate, as an alternative scenario, the presence of gas-phase water could also reduce the \ce{HCO^+} abundance \citep{2018A&A...613A..29V, 2021A&A...646A...3L, 2021A&A...650A.180N}. HD~100546 is one of the only disks that has a detection of \ce{H_2O} gas with \textit{Herschel} \citep{2021A&A...648A..24V, 2022A&A...665A..45P} and water ice in the surface layers \citep{2016ApJ...821....2H}. 

\subsubsection{UV field} 

The column density ratio of CN to HCN has been long suggested as a probe of the strength of the UV field impinging onto protoplanetary disks \citep{2012A&A...537A..60C}. Due to different photo-dissociation cross sections of CN and HCN, CN is predicted to survive in less dense and more UV irradiated regions therefore in a higher vertical layer in protoplanetary disks than HCN \citep[e.g.,][]{2018A&A...616A..19A}. \citet{2021ApJS..257...11B} use high angular resolution observations of HCN and CN to explore photochemistry across the HD~163296 and MWC~480 disks. They find that N(CN)/N(HCN) increases as a function of radius and the value of this ratio is always $>$10. The resulting HCN and CN column density profiles from \citet{2021ApJS..257...11B} are shown in Figure~\ref{fig:comparison} along with our values for HD~100546. It is immediately apparent that CN is under-abundant across the whole of the HD~100546 disk compared to the others by more than an order of magnitude. The CN abundance in disks has been shown to be sensitive to the UV field and the degree of flaring in the protoplanetary disk \citep{2018A&A...609A..93C}. All three stars have similar UV luminosities but HD~100546 is slightly brighter, and this would be predicted to give a higher CN abundance. Therefore, it could be that the disk structure is responsible for the differences in CN abundance \citep{2012A&A...544A..78M}. The CN and HCN in the HD~163296 disk have been shown to originate from a layer in the disk at Z/R$\approx$0.2 but for MWC~480 this these molecules are emitting from Z/R$<$0.1 \citep{2021ApJS..257....4L, 2023A&A...669A.126P}.
%
%
Our analysis in Figure~\ref{fig:ellipses} indicates these species come from close to the midplane in the HD~100546 disk with Z/R$<$0.1. Therefore, a molecular layer closer to the midplane due to a shallow disk and/or a higher UV flux pushing the molecular later deeper into the disk could explain the lack of CN in the HD~100546. In comparison, in the inner disk the HCN column density is significantly lower than the two other disks but the HCN in the HD~100546 outer ring is within a factor of a few. The lack of HCN (and CN) in the inner disk is at least partly due to the $\approx$15~au gas cavity in HD~100546 which is not present in the other two disks. Overall we find that N(CN)/N(HCN) over the HD~100546 disk ranges from $10^{-1}$ to 10 from 10 to 350~au which is different to the other disks observed so far but is broadly consistent with the thermochemical models from \citet[][]{2014ApJ...797..113S} which have no volatile depletion of C and O. 

\subsubsection{C/O}
Individual molecules such as \ce{C_2H} and pairs of molecules, e.g., CS/SO and CN/NO, can be used to unravel the underlying elemental C/O in disk gas \citep[e.g.,][]{2018A&A...617A..28S,2019A&A...631A..69M,2019ApJ...886...86L}. In the HD~163296 and MWC~480 disks there is evidence for an elevated C/O of $\approx$2 in the disk warm molecular layer traced by \ce{C_2H} \citep{2021ApJS..257....7B}. As shown in Figure~\ref{fig:comparison}, 
 \ce{C_2H} is abundant out to $\approx$150~au in these disks.
    In the HD~100546 disk, the peak \ce{C_2H} column density is $\approx100\times$ lower that of HD~163296 and MWC~480.
    This difference can attributed to the lack of CO freeze-out and volatile depletion in HD~100546 compared to these other two disks despite the extreme dust evolution. \citet{2016A&A...592A..83K} show that the C/H and O/H in HD~100546 are consistent with interstellar abundances and are at most modestly depleted $<10\times$. Whereas, the models from \citet{2021ApJS..257....7B} require a depletion of C/H of 4-10$\times$ and O/H 20-50$\times$ to reproduce the high column densities observed towards HD~163196 and MWC~480.
    The second ring of \ce{C_2H} in HD~100546 peaks just beyond the second dust ring and, as proposed by \citet{2021ApJ...910....3B}, may trace gas with a specific balance between UV irradiation and density that favours the formation of \ce{C_2H}. Interestingly though, the \ce{C_2H} is not bright within the dust gap as proposed by \citet{2021ApJ...910....3B}. 
    
    In comparison, the CS column density in HD~100546 is within a factor of a few compared to MWC~480 \citep[see Figure~\ref{fig:comparison}][note that a radial CS column density profile for HD 163296 was not provided the MAPS data release]{2021ApJS..257...12L}. This may indicate that the elevated C/O traced by \ce{C_2H} in the MAPS disks is not necessarily traced by the CS. 
    One mechanism to enhance the volatile carbon in the surface layers of protoplanetary disk is via the destruction of refractory carbon \citep{2017ApJ...845...13A,2021ApJ...910....3B}. CS is predicted to come from a lower layer in the disk than \ce{C_2H} \citep{2018A&A...616A..19A} therefore if this is a process occurring in the higher disk layers it might not be traced as effectively with CS. But, the CS abundance is limited by the volatile sulphur abundance which is significantly lower than that of carbon. We do though see clear evidence for radial C/O variations in the HD~100546 disk traced by the SO and CS where SO is abundant in the cavity (see Figure~\ref{fig:hd100546_sulphur}) as also reported for HD~169142 \citep{Booth2023}. Our SO peak column density in HD~100546 is $\approx100\times$ higher than the disk-averaged upper-limit in the MAPS Herbig disks of $<3\times10^{12}$~cm$^{-2}$. The weak SO emission in the outer disk of HD~100546 has a column density of $\approx3\times10^{12}$~cm$^{-2}$ so if this reservoir of cold SO is present in HD~163296 and MWC~480 it may just be below the detection limit of the MAPS data. 
    
    In HD~100546, NO is the nitrogen-bearing molecule with the highest column density we have detected. NO is not typically targeted in disk observations but, the high NO/CN and overall low CN abundance in HD~100546 disk may also be related to the lack of volatile depletion \citep{2019ApJ...886...86L, 2023arXiv230300768L}. Across the HD~100546 disk the N(CN)/N(NO) varies from $10^{-1}$ to 1. 
    Detailed thermochemical modelling from Leemker et al. (2023, in prep) will determine more quantitative constraints on the C/O ratio across the HD~100546 to better place it in context with other Class II disks.

\subsubsection{Large organics}

The organic molecule \ce{H_2CO} is ubiquitous in protoplanetary disks \citep[e.g.,][]{2020ApJ...890..142P}. The \ce{H_2CO} detected in disks can form in the gas phase via neutral-neutral reactions, e.g., \ce{CH_3} + O, or on the surfaces of icy dust grains via CO ice hydrogenation \citep[e.g.,][]{2002ApJ...571L.173W, 2015ApJ...809L..25L}. \ce{H_2CO} is detected in HD~100546 with a similar column density across the disk to HD~163296 and MWC~480, as shown in Figure~\ref{fig:comparison}, despite the lack of significant CO freeze-out in HD~100546 (as also seen for HD~169142; \citealt{Booth2023}). This could hint at the gas-phase formation of \ce{H_2CO} rather an CO ice hydrogenation as being the dominant in-situ source of \ce{H_2CO} in these disks, but if these disks begin their evolution with inherited \ce{CH_3OH} rich ices then \ce{H_2CO} is likely also present on the grains in these disks regardless of the freeze-out of CO.   

Since \ce{H_2CO} and \ce{CH_3OH} have a common formation pathway on grain surfaces via CO ice \citep[e.g.,][]{2002ApJ...571L.173W, 2009A&A...505..629F, 2022ApJ...931L..33S} the former has been proposed as a tracer of the underlying complexity in disks. Unfortunately, the general lack of \ce{CH_3OH} in \ce{H_2CO}-rich sources shows that is not that simple. In HD~100546 we find a peak abundance ratio of \ce{CH_3OH}/\ce{H_2CO} of $18\pm4$ in the inner disk at 100~K and $1.1\pm0.6$ in the outer disk at 30~K. This is a few times lower than the HD~169142 \ce{CH_3OH}/\ce{H_2CO} ratio of 100 of the inner disk and the upper-limit of $\approx$3 for the outer disk \citep{Booth2023}. 
In HD~163296 and MWC~480 the upper limits on the disk averaged \ce{CH_3OH} column densities are $<2\times10^{12}$ and $<8\times10^{12}$~cm$^{-2}$ so if the same reservoir of, particularly warm thermally desorbed, \ce{CH_3OH} was present in these sources it should have been detected. 
The lack of COMs emission could be attributed to disk structure instead of disk chemistry. In HD~100546 the COMs can desorb at the edge of the millimetre dust cavity where as the millimetre dust is optically thick within the \ce{H_2O} snowline locations in HD~163296 and MWC~480 disks \citep{2021ApJS..257....5Z,2021ApJS..257...14S}. 
We also detect \ce{CH_3OCHO} in HD~100546 at a high relative abundance to \ce{CH_3OH}. This larger molecule is also present in IRS~48 but,  unlike in the IRS~48 disk \citep{2022A&A...659A..29B}, another COM \ce{CH_3OCH_3} is not detected in HD~100546 with a \ce{CH_3OCHO}/\ce{CH_3OCH_3} column density ratio $>$7.  A more detailed discussion on the COMs reservoir in the HD~100546 disk will follow in \citet{Booth2023_irs48} and Evans et al. (in prep.). 


The sulphur-bearing equivalent of \ce{H_2CO}, \ce{H_2CS}, is detected in HD~100546 and follows the CS (also found for MWC~480 and HD~169142; \citealt{2019ApJ...876...72L, 2021ApJS..257...12L, Booth2023}) and not the sublimating SO in the inner disk. This indicates that the \ce{H_2CS} is likely forming in the gas phase rather than having a significant abundance on the grains. The column density ratio observed towards HD~100546 is within a factor of a few of that measured in the other two Herbig disks, HD~169142 
and MWC~480. OCS is not detected in our data and we place an upper-limit on the column density of 
$<$10$^{12}$~cm$\mathrm{^{-2}}$ at 100~K, 
which corresponds to $<$1\% of the SO column density and thus we can conclude that OCS not a significant source of gaseous volatile sulphur in this disk. 

We do not detect \ce{CH_3CN}, \ce{c-C_3H_2} or \ce{HC_3N} in our observations of the HD~100546 disk but the lines covered do have higher upper energy levels than those targeted in MAPS \citep{2021ApJS..257....9I}. 
Using the upper limits on the disk-integrated fluxes (from Table~\ref{tab:A3}) we calculate disk-integrated column density upper limits at 50~K within a radius of 100~au.
This results in a column density 
$<$2$\times$10$^{13}$
~cm$\mathrm{^{-2}}$ 
for \ce{CH_3CN}, 
$<$1$\times$10$^{14}$~cm$\mathrm{^{-2}}$ for \ce{c-C_3H_2} and
$<$6$\times$10$^{12}$~cm$\mathrm{^{-2}}$ for \ce{HC_3N}.
Compared to \citet{2021ApJS..257....9I} these upper limits are not sensitive enough compared to the \ce{HCN} and \ce{C_2H} column densities to detect the same level of complexity as seen in the HD~163296 and MWC~480 disks. Additionally, the abundance of these species is expected to be enhanced in high C/O environments therefore, we might expect them to be less abundant in HD~100546 relative to HCN and 
\ce{C_2H} \citep{2023NatAs...7...49C}. 
The lack of these species could also indicate that the small grains within the dust cavity are not hot enough to trigger carbon grain sublimation which would result in the in-situ gas-phase formation of of hydrocarbons and nitriles \citep{2020ApJ...897L..38V}.

\section{Conclusion} \label{sec:conclusion}

This paper presents an ALMA molecular line survey towards the disk around the Herbig Ae star HD~100546. We detect 19 different molecular species and our main results are as follows: 

\begin{itemize}
    \item This work reports the first detections of \ce{H_2^{13}CO} and \ce{^{34}SO} in protoplanetary disks. We also robustly detect \ce{CH_3OCHO} along with \ce{SO_2} and \ce{NO} showing that these molecules are detectable in Class II disks. 
    
    \item The molecular emission from the HD~100546 disk shows clear radial sub-structures that peak just beyond the the outer millimetre dust ring at $\approx$200~au and these emission rings appear to be flat. This spatial coincidence is in contrast with what is generally seen in other sources. A similarly clear gap is not seen in the CO isotopologues indicating that a deep gas (\ce{H_2}) gap cannot solely explain the presence of these molecular rings. 
    
    \item There is a consistent azimuthal brightness asymmetry in the outer HD~100546 molecular rings that is most prominent in the radicals \ce{C_2H}, CN and CS. This could be due to shadowing from the inner dust disk leading to an azimuthal temperature gradient and/or varying incident UV irradiation. There is also asymmetric SO emission originating from within the central dust cavity which may be linked to ongoing giant planet formation, as proposed in \citet{2023A&A...669A..53B}.  

    \item The fractional abundance of \ce{HCO^+} relative to CO in HD~100546 is $\approx10^{-7}$. This is 1-3 orders of magnitude lower than the ranges reported for the other Herbig Ae disks HD~163296 and MWC~480, but is consistent with the HD~142527 disk. This may be explained by a different ionization structure in transition and non-transitional disks. The presence of gas-phase \ce{H_2O} in the HD~100546 disk will also contribute to the lack of \ce{HCO^+}. 

    \item The low CN abundance relative to CO and HCN in the HD~100546 disk is unique when compared to other sources. The stellar UV flux from HD~100546 is similar to that of HD~163296 and MWC~480 but the CN is $\approx10-100\times$ less abundant across the disk. 
    If the CN emitting layer in HD~100546 is close to the midplane due to a flat or self-shadowed disk, as supported by the morphology of the outer ring, the CN abundance could be expected to be low. Chemical modelling is needed to test this hypothesis.  

    \item We detect multiple species that can be used to trace the underlying C/O ratio in the disk gas. The \ce{CS}/\ce{SO} clearly trace the transition over the \ce{H_2O} snowline in the inner disk. 
    The \ce{C_2H} column density is $\times$100 lower than other Herbig disks of a similar mass indicating less oxygen depletion in this system and suggesting a low C/O. 
    The \ce{CS}, \ce{C_2H} and \ce{CN} have similar radial emission profiles and could trace rings in the disk where C/O$\approx$1. The relationship between CN and NO, the SO and CS, and the \ce{C_2H} in the outer ring in the context of radial C/O variations requires further investigation.    

    \item Both \ce{CH_3OH} and \ce{CH_3OCHO} are detected in the HD~100546 disk indicating a hot corino/core-like chemistry may be active in the inner region of disks. In comparison to other disks, \ce{CH_3CN}, \ce{HC_3N} and \ce{c-C_3H_2} are all undetected in our data but, this can primarily be attributed to sensitivity limits of the data. 
    
\end{itemize}

The chemical diversity and richness of the spectra we present here motivate the need for more uniform surveys of protoplanetary disks around Herbig Ae stars. These class of disks are typically larger and brighter than T-Tauri disks and the sublimation of ices in these disks makes them excellent observational laboratories to unravel disk chemistry at the time of planet formation. It is particularly important to understand the ongoing chemistry in these disks as they are the precursors to the star, debris disk and giant-planet systems like HR~8799 and Beta Pictoris \citep{2008Sci...322.1348M, 2009A&A...493L..21L}.

\begin{acknowledgements}
This paper makes use of the following ALMA data: 2021.1.00738S. We acknowledge assistance from Allegro, the European ALMA Regional Centre node in the Netherlands. ALMA is a partnership of ESO (representing its member states), NSF (USA) and NINS (Japan), together with NRC (Canada), MOST and ASIAA (Taiwan), and KASI (Republic of Korea), in cooperation with the Republic of Chile. The Joint ALMA Observatory is operated by ESO, AUI/NRAO and NAOJ. 
This work has used the following additional software packages that have not been referred to in the main text: Astropy, IPython, Jupyter, Matplotlib and NumPy \citep{Astropy,IPython,Jupyter,Matplotlib,NumPy}.
Astrochemistry in Leiden is supported by funding from the European Research Council (ERC) under the European Union’s Horizon 2020 research and innovation programme (grant agreement No. 101019751 MOLDISK).
A.S.B. is supported by a Clay Postdoctoral Fellowship from the Smithsonian Astrophysical Observatory. 
M.L. acknowledges support from grant 618.000.001 from the Dutch Research Council (NWO).
J.I.D. acknowledges support from an STFC Ernest Rutherford Fellowship (ST/W004119/1) and a University Academic Fellowship from the University of Leeds.
M.T. acknowledges support from the ERC grant 101019751 MOLDISK. 
C.W.~acknowledges financial support from the University of Leeds, the Science and Technology Facilities Council, and UK Research and Innovation (grant numbers ST/X001016/1 and MR/T040726/1).
L.E. acknowledges financial support from the Science and Technology Facilities Council (grant number ST/T000287/1).
S.N.~is grateful for support from RIKEN Special Postdoctoral Researcher Program (Fellowships), Grants-in-Aid for JSPS (Japan Society for the Promotion of Science) Fellows Grant Number JP23KJ0329, and MEXT/JSPS Grants-in-Aid for Scientific Research (KAKENHI) Grant Numbers JP 18H05441, JP20K22376, JP20H05845, JP20H05847, JP23K13155, and JP23H05441. 
Support for C.J.L. was provided by NASA through the NASA Hubble Fellowship grant No. HST-HF2-51535.001-A awarded by the Space Telescope Science Institute, which is operated by the Association of Universities for Research in Astronomy, Inc., for NASA, under contract NAS5-26555. 
\end{acknowledgements}

\bibliography{sample631}
\bibliographystyle{aasjournal}

\newpage
\appendix
\section{Observational set-up}

           
           



\begin{table}[h!]
    \centering
        \caption{Execution block details for ALMA program 2021.100738.S targeting HD~100546.}
    \begin{tabular}{c c c c c c c c c c} \hline \hline
        Setting & Date & No. Antenna & On Source Time& Baselines & Mean PVW & MRS &Phase & Flux/Bandpass\\
                 &     &   & (mins)  &  (m) & (mm)& (") & Calibrator & Calibrator \\ \hline
       A  &  05-01-2022   & 42  & 62 & 14.9-783.1 & 0.5 & 3.5 & J1147-6753 & J142-4206 \\
          &  27-05-2022   &  46 & 62 & 15.1-783.5 & 0.8 & 3.9& J1147-6753 & J142-4206 \\ 
          &   09-07-2022    & 41 & 62  & 15.1-783.5 & 0.2 & 3.4&J1147-6753 & J142-4206 \\ 
         &  10-07-2022    & 41  & 62  & 15.1-783.5 & 0.1 & 3.4 & J1147-6753 & J142-4206 \\
        B &  05-01-2022   &  42 & 53 & 14.9-783.1 &0.5 & 3.5 & J1147-6753 & J142-4206 \\
          &  06-01-2022   & 45 & 53 & 14.9-976.6& 1.0& 3.5& J1147-6753 & J142-4206 \\
         &    10-07-2022  &41 & 53 & 15.1-1213.4& 0.5& 3.7 &J1147-6753 & J142-4206\\
     &    13-07-2022   & 43 & 53 & 15.1-1213.4& 0.2 & 3.2& J1147-6753 & J142-4206\\
              &  14-07-2022   & 44 & 53  & 15.1-1301.6& 0.3& 2.8& J1147-6753 & J142-4206 \\ \hline \\
        \end{tabular}
    \label{tab:A3}
\end{table}


\newpage
\section{Molecular Data}

\begin{table*}[h!]
   \caption{Molecular data of the transitions presented in this paper. This covers all of the molecules detected in the disk and particular non-detections of interest but not all of the transitions covered/detected for these species. All data are taken from CDMS \citep{2016JMoSp.327...95E} except for \ce{C^{17}O}, \ce{C_2H}, \ce{CH_3OCHO} and \ce{CH_3OCH_3} which are from JPL \citep{1998JQSRT..60..883P}.} 
    \begin{tabular}{c c c c c c c}
    \hline
    Molecule & Transition & Frequency (GHz) & E$_{\mathrm{up}}$ (K )& log10${\mathrm{(A_{ul}}}$) & $\mathrm{g_{u}}$ & Detection \\  
    \hline \hline

    \ce{^{12}CO}     & $J=3-2$ & 345.7959899  & 33.2 & -5.6027  & 7   & \cmark \\
    \ce{C^{17}O}     & $J=3-2$ & 337.0611298  & 32.7 & -5.6344  & 7   & \cmark \\
    \ce{HCO^+}       & $J=4-3$ & 356.7342230  & 42.8 & -2.4471  & 9   & \cmark \\
    \ce{HC^{18}O^+}  & $J=4-3$ & 340.6306916  & 41.8 & -2.5310  & 12  &  -
 \\
    \ce{HCN}         & $J=4-3$ & 354.5054779  & 42.5 & -2.6860  & 27  & \cmark \\
    \ce{H^{13}CN}    & $J=4-3$ & 345.3397693  & 41.4 & -2.7216  & 27  & \cmark \\
    \ce{HC^{15}N}    & $J=4-3$ & 344.2001089  & 41.3 & -2.7258  & 9   & \cmark\\

               CN      & $J=7/2-5/2, F=7/2-5/2$ & 340.2477700 & 32.7 &	-3.3839  & 10 & \cmark \\
                     & $J=7/2-5/2, F=7/2-5/2$ & 340.2477700 & 32.7 & -3.4206  & 8  & \cmark \\
                     & $J=7/2-5/2, F=5/2-3/2$ & 340.2485440 & 32.7 & -3.4347  & 6   & \cmark \\
                    

    \ce{NO}                   & $J=7/2-5/2$,$\Omega=1/2- F=9/2-7/2$     & 351.0435240   &  36.1 & -5.2649  &  10    & \cmark \\
                     & $J=7/2-5/2$,$\Omega=1/2- F=7/2-5/2$     & 351.0517050   &  36.1 & -5.2662 &  8   & \cmark \\
                     & $J=7/2-5/2$, $\Omega=1/2- F=7/2-5/2$     & 351.0517050   &  36.1 & -5.3161 &  6   & \cmark \\

    \ce{HC_3N}        & $J=38-37$  & 345.6090100 & 323.5 &  -2.4812   & 77 & -\\
                      &  $J=39-38$	 & 354.6974631 & 340.5 &  -2.4473 & 79  & - \\

    \ce{CH_3CN}     &   $J=19_0-18_0 $& 349.4536999 & 167.7  & -2.5909  & 78 & - \\

                \ce{C_2H}       & $J=9/2-7/2,F=5-4$  & 349.3374558  & 41.9 & -3.7247 & 11   & \cmark \\
                     & $J=9/2-7/2,F=4-3$  & 349.3387284  & 41.9 & -3.7349 & 9   & \cmark \\

    \ce{c-C_3H_2}     &  $J=10_{(0,10)}-9_{(1,9)}$ & 351.7815780 & 96.5& -2.6125 & 63 & - \\
                       &  $J=10_{(0,10)}-9_{(1,9)}$ & 351.7815780 & 96.5  & -2.6126 & 21 & - \\

    \ce{CS}          & $J=7-6$  & 342.8828503 & 65.8 & -3.0774  & 15  & \cmark \\
    \ce{C^{34}S}     & $J=7-6$  & 337.3964590 & 50.2 & -3.1180  & 15  & \cmark\\

    \ce{H_2CS}       & $J=10_{(1,10)}-9_{(1,9)}$ & 338.0831953  & 102.4     & 	-3.1995  & 63   & \cmark \\ 
                     
    \ce{SO}          & $J=3_3-3_2$  & 339.3414590 & 25.5 & -4.8372   & 7  &  - \\
                     & $J=7_8-6_7$  & 340.7141550 & 81.2 & -3.3023   & 15 &  \cmark \\ 
                     & $J=8_8-7_7$  & 344.3106120 & 87.5 & -3.2852   & 17 &  \cmark \\

    \ce{^{34}SO}     & $J=8_8-7_7$  & 337.5801467 & 77.3 & -3.3109 & 17 &  -     \\
                     & $J=9_8-8_7$  & 339.8572694 & 86.1 & -3.2944 & 19 &  \cmark \\
                      
    \ce{SO_2}        &  $J=6_{(4,2)}-6_{(3,3)}$ &  357.9258478 & 58.6 & -3.5845  & 13 &  \cmark \\

    \ce{OCS}         & $J=28-27$  & 340.4492733 & 237.0 &  	-3.9378 & 57 &   - \\
                     & $J=29-28$  & 352.5995703 & 253.9 &  	-3.8918 & 59 &   -\\

    \ce{H_2CO}       & $J=5_{(1,5)}-4_{(1,4)}$ & 351.7686450  & 62.5 & 	-2.9201 & 33 &  \cmark\\
    
    \ce{H_2^{13}CO}  & $J=5_{(1,5)}-4_{(1,4)}$ & 343.3257130  & 61.3 &  -2.9517    & 33 &  \cmark \\

    \ce{CH_3OH}      & $J=7_{0}-6_{0}$ &  338.4086980& 65.0 &  -3.7691 & 60 & \cmark \\
    


        \ce{CH_3OCHO}    & $J=32_{(2,31)}-31_{(2,30)}$ & 344.0297653 &276.1 & -3.2099& 65 &  \cmark \\
 & $J=32_{(1,32)}-31_{(1,31)}$  &  344.0297645&276.1&-3.2099 & 65 &  \cmark \\
  &$J=32_{(0,32)}-31_{(0,31)}$  & 344.0295703 &276.1 & -3.2099& 65 &  \cmark \\
   & $J=32_{(1,32)}-31_{(1,31)}$  & 344.0295694 &276.1&-3.2099 & 65 &  \cmark \\


    \ce{CH_3OCH_3}   & $J=19_{(0,19)}-18_{(1,18)}$ AE & 342.6080601 &167.1& -3.2816 & 117 &  - \\ 
 &  $J=19_{(0,19)}-18_{(1,18)}$ EA &  342.6080602& 167.1&  -3.2817& 78 & -\\ 
  & $J=19_{(0,19)}-18_{(1,18)}$ EE & 342.6081188 &167.1 & -3.2816 & 312 &- \\ 
   & $J=19_{(0,19)}-18_{(1,18)}$ AA & 342.6081774 &167.1 & -3.2816 & 195 &- \\



    \hline
    \end{tabular}
    \label{tab:lines}
\end{table*}

\newpage
\section{Image properties}

\begin{table*}[h!]
    \caption{Properties of the line images for HD~100546.}
    \centering
    \begin{tabular}{c c c c c c c c c}
    \hline
    Molecule & Transition & robust & Beam & rms & Peak & Int. Flux \\  
    &  &  & (\farcs $\times$ \farcs ($^{\circ}$)) & (mJy beam $^{-1}$) &  (mJy beam $^{-1}$) & (mJy km s$^{-1}$) \\
    \hline \hline
    \ce{^{12}CO}    & $J=3-2$  & 0.5 &  0.35$\times$0.23~(59.7) &  0.76 & 1128.66  & 190464.0$\pm$28.0\\
    \ce{C^{17}O}    & $J=3-2$  & 0.5 &  0.39$\times$0.31~(28.5) &  0.86 & 216.49   &   5693.0$\pm$26.0\\
    \ce{HCO^+}      & $J=4-3$  & 0.5 &  0.34$\times$0.23~(59.8)	& 0.88	& 48.81    & 5952.0$\pm$33.0 \\
    
    \ce{HC^{18}O^+} & $J=4-3$  & 2.0 &  0.37$\times$0.29~(28.8)&0.78& - &$<$33.0 \\
    \ce{HCN}        & $J=4-3$  & 0.5 &  0.34$\times$0.23~(59.7)&0.81& 72.58 &3699.0$\pm$30.0\\
    \ce{H^{13}CN}   & $J=4-3$  & 0.5 &  0.35$\times$0.23~(59.7)&0.72& 6.12 &172.0$\pm$26.0\\
    \ce{HC^{15}N}   & $J=4-3$  & 0.5 &  0.35$\times$0.24~(59.3)&0.78& 8.17 &134.0$\pm$29.0\\
    \ce{CN}         & $N=4-3$  & 0.5 &  0.38$\times$0.30~(29.5)&0.83&28.77&2732.0$\pm$26.0\\
    
    
    \ce{NO}         &  $J=7/2-5/2$ & 2.0    & 0.47$\times$0.37~(30.8)  &0.73&  3.75 &134.0$\pm$27.0\\
    \ce{HC_3N}    & $J=38-37$  &  2.0       & 0.47$\times$0.33~(58.3)  &0.67   & -  &  $<$23.0 \\
                    & $J=39-38$   & 2.0     & 0.46$\times$0.32~(58.5)  &0.69   & -  &  $<$25.0 \\

    \ce{CH_3CN}   & $J=19_0=18_0$& 2.0      & 0.46$\times$0.37~(30.7)   & 0.68  &  - & $<$24.0\\

    \ce{C_2H}       & $N=4-3$  & 0.5 &  0.37$\times$0.30~(26.7)&0.80 & 20.88&1601.0$\pm$25.0\\
      
    \ce{c-C_3H_2}   & $J=10_{(0,10)}-9_{(1,9)}$& 2.0 & 0.47$\times$0.37~(30.8)   &  0.70 & - & $<$27.0  \\
    
    \ce{CS}         & $J=7-6$ & 0.5 &  0.35$\times$0.24~(59.4)&0.77&30.38&932.0$\pm$28.0\\
    \ce{C^{34}S}    & $J=7-6 $& 2.0 &  0.48$\times$0.39~(31.9)&0.74& 7.00 &86.0$\pm$18.0\\
    \ce{H_2CS}      &$J=10_{(1,10)}-9_{(1,9)}$ & 2.0 &   0.38$\times$0.30~(27.7)&0.85& 11.69 &128.0$\pm$27.0\\

    \ce{SO}         & $J=7_8-6_7$ & 0.5 &    0.37$\times$0.30~(28.8)&0.82 &25.79&343.0$\pm$26.0\\
                    & $J=8_8-7_7$ & 0.5 &    0.35$\times$0.23~(59.6)&0.75& 23.69 &281.0$\pm$27.0\\
                    &$J=3_3-3_2$ & 0.5 & 0.38$\times$0.30~(28.8)   & 0.76  & -  &  $<$32.0 \\

    \ce{^{34}SO}    & $J=8_8-7_7$ & 0.5 &    0.38$\times$0.30~(28.8)&0.77& 3.49& 32.0$\pm$10.0\\
                    & $J=7_8-6_7$ & 0.5 &    0.38$\times$0.30~(27.8)&0.82& - & $<$33.0\\

    \ce{SO_2}       & $J=6_{4,2}-6_{3,3}$ & 2.0 &  0.47$\times$0.32~(60.8)&0.78& 5.39 &98.0$\pm$21.0\\
    
    \ce{OCS}   & $J=28-27$ & 2.0 &  0.48$\times$0.37~(32.8)  &  0.69 &  -& $<$21.0 \\
        & $J=27-26$        & 2.0 &   0.46$\times$0.37~(32.8) &  0.97 &  -& $<$33.0 \\

    \ce{H_2CO}      & $J=5_{(1,5)}-4_{(1,4)}$& 0.5 &  0.37$\times$0.30~(27.1)&0.94& 55.27 &3065.0$\pm$30.0\\

    \ce{H_2^{13}CO} & $J=5_{(1,5)}-4_{(1,4)}$& 2.0 &  0.47$\times$0.33~(58.2)&0.79& 3.88 &66.0$\pm$21.0\\

    \ce{CH_3OH}     & $J=7_{0}-6_{0}$ & 0.5 &  0.38$\times$0.30~(28.3)&0.80& 6.92 &39.0$\pm$12.0\\
    
    \ce{CH_3OCHO}   & $J=31-30$ & 0.5 &  0.35$\times$0.24~(59.3)&0.76& 8.47 &82.0$\pm$12.0\\
    
    \ce{CH_3OCH_3}&$J=19-18$ & 0.5   & 0.36$\times$0.24~(59.4)&0.83& - & $<$39.0\\   

    \hline
    \end{tabular}
    \label{tab:hd100546_images}
\end{table*}




\newpage
\section{Weak-line detections}

\begin{figure*}[h!]
\centering
    \includegraphics[width=0.95\hsize]{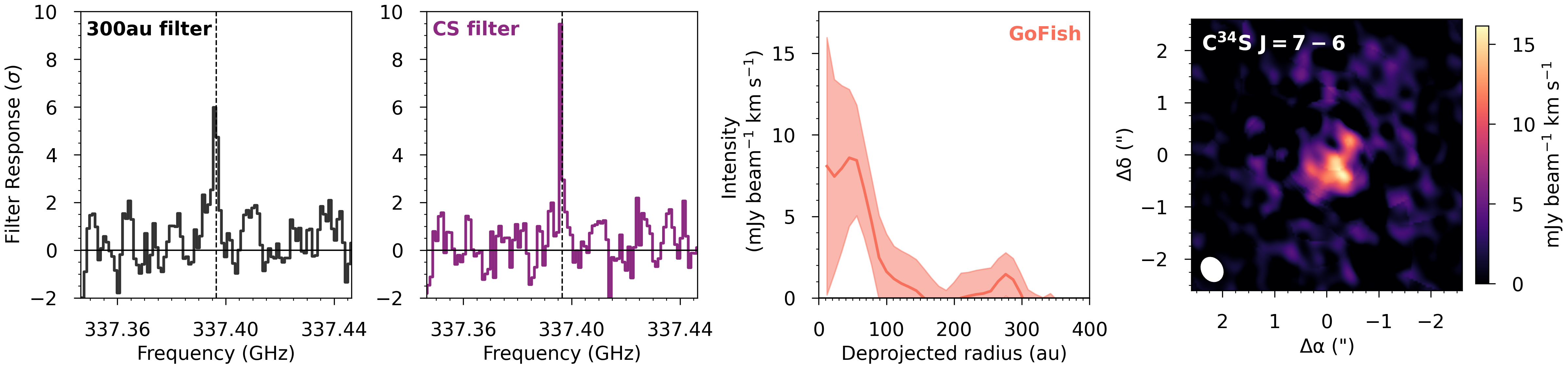}
    \includegraphics[width=0.95\hsize]{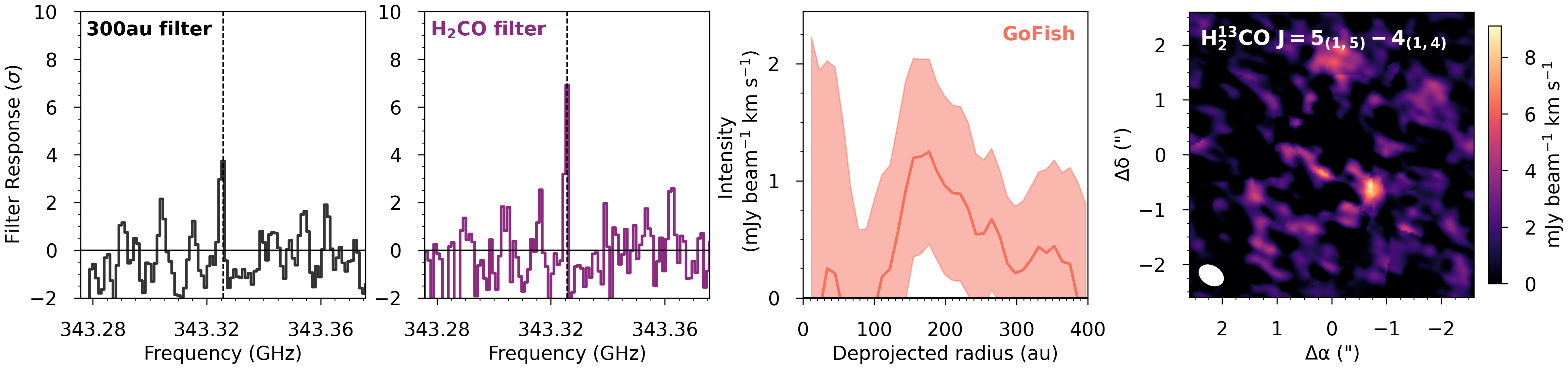}
    \includegraphics[width=0.95\hsize]{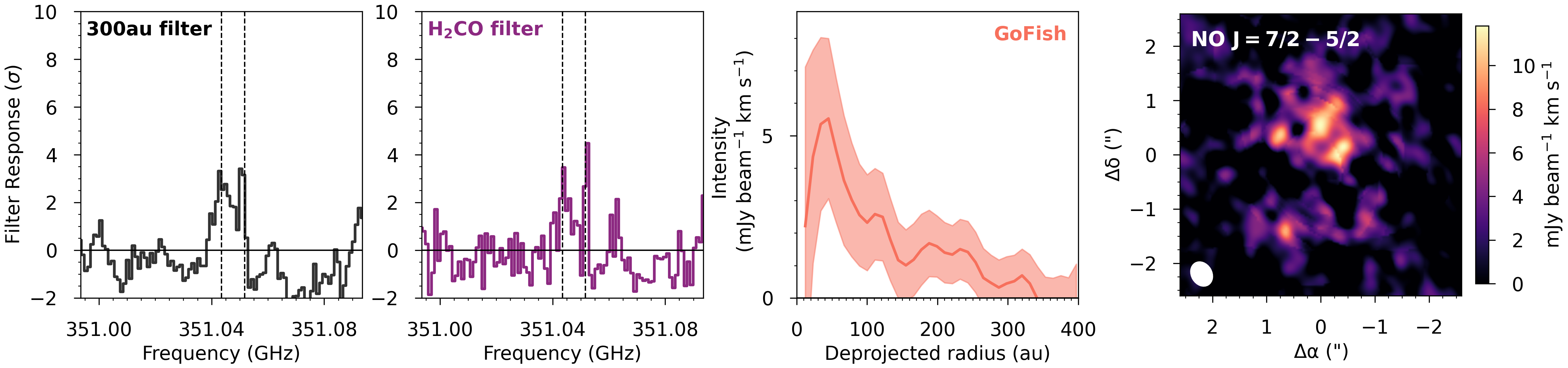}
    \includegraphics[width=0.95\hsize]{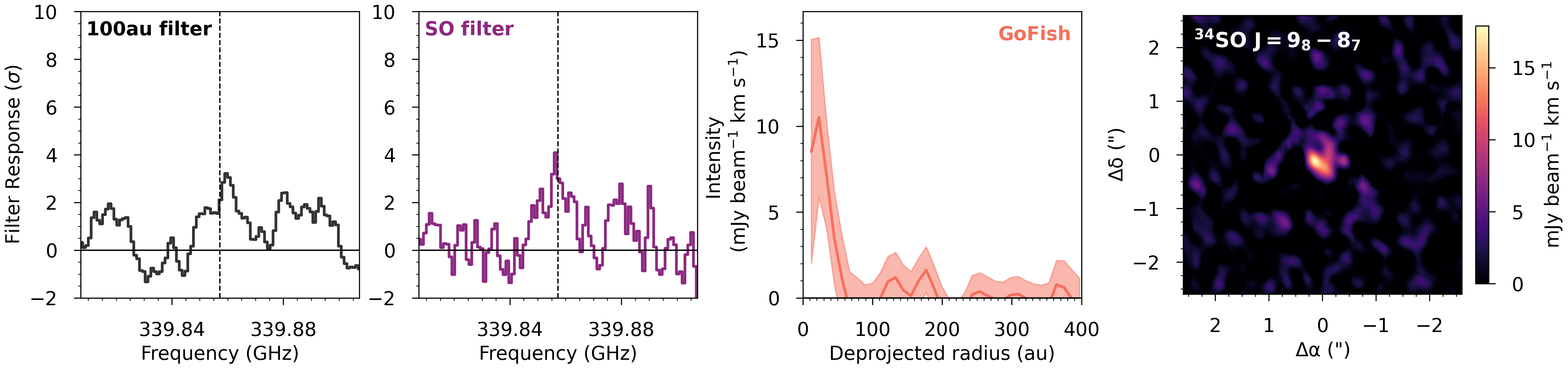}
    \includegraphics[width=0.95\hsize]{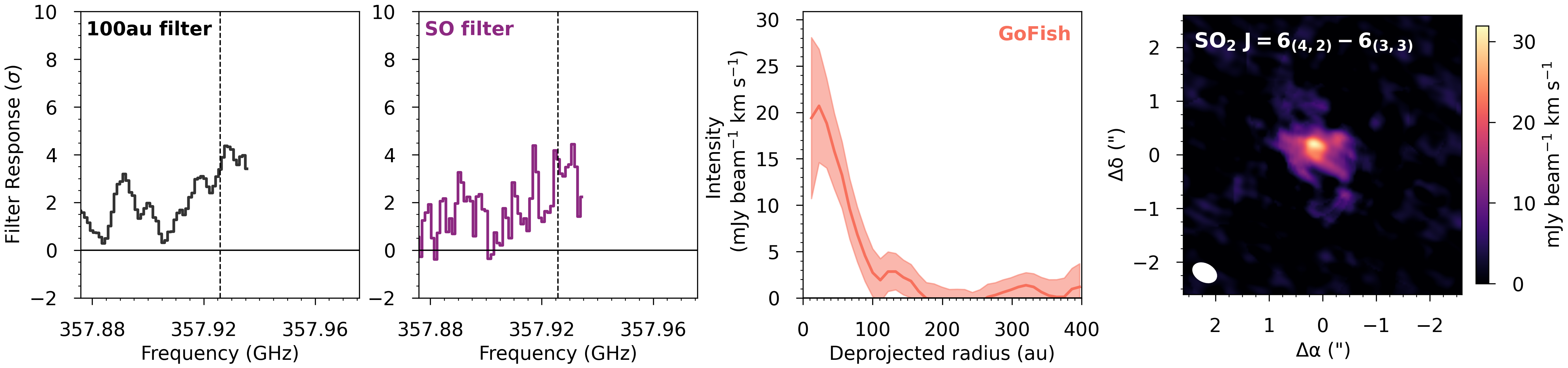}
    \caption{Matched filter responses with Keplerian models and strong lines as filters alongside GoFish radial profiles and Keplerian masked integrated intensity maps for \ce{C^{34}S}, \ce{H_{2} ^{13}CO}, NO, \ce{^{34}SO} and \ce{SO_2} lines in the HD~100546 disk.}
    \label{fig:weaklines}
\end{figure*}

\begin{figure*}
\centering
    \includegraphics[width=0.95\hsize]{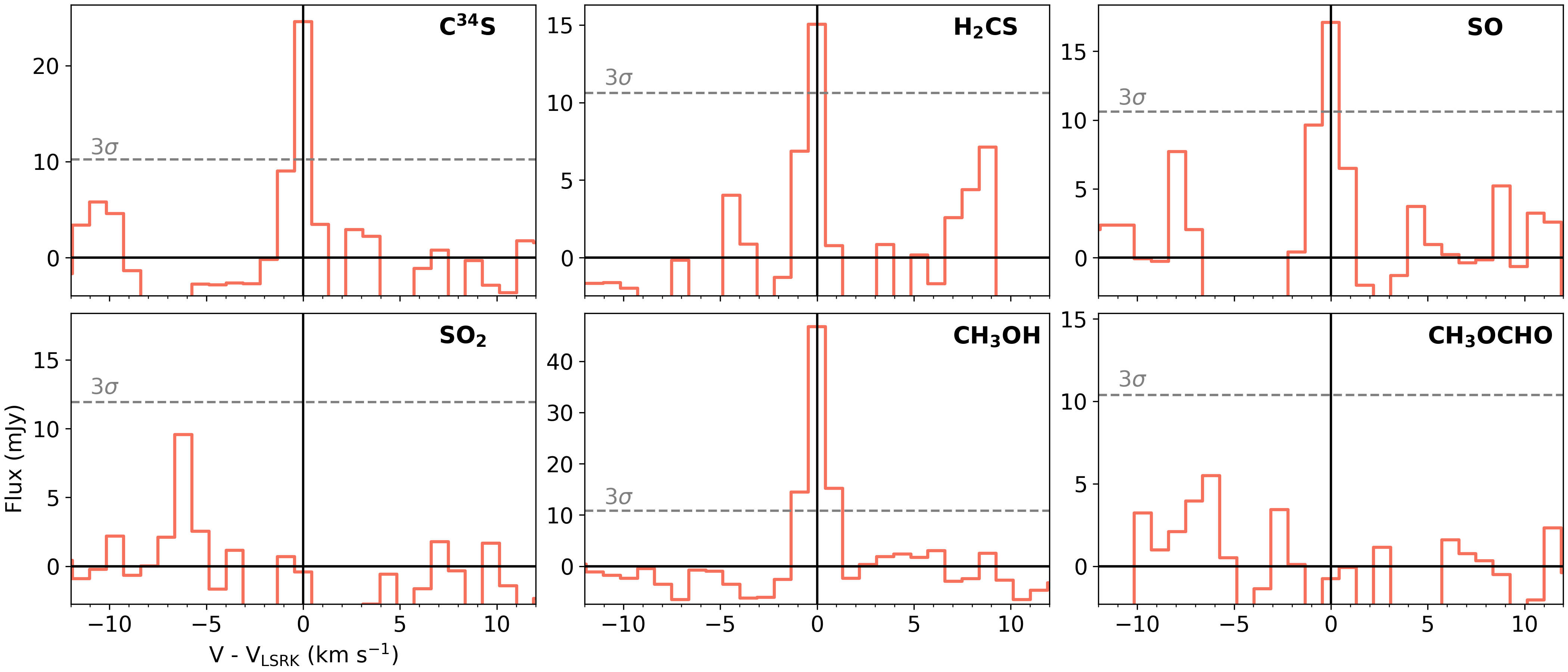}
    \caption{Spectra generated using \textit{GoFish} stacked over a 180 to 250~au annulus \citep{GoFish}.}
        \label{fig:gofish}
\end{figure*}

\end{document}